\documentclass[8pt]{article}

\usepackage{amsmath,amsxtra,amsthm,amssymb,xr,mathtools}
\usepackage{graphicx}
\usepackage[latin1]{inputenc}
\usepackage{hyperref}
\usepackage[all]{xy}
\usepackage{xcolor}
\usepackage{stmaryrd}
\usepackage{rotating}
\usepackage{dsfont}
\usepackage{subcaption}
\usepackage{float}
\usepackage{diagbox}
\usepackage{cancel}
\usepackage{adjustbox}

\newcommand{\bit}{\begin{itemize}}
\newcommand{\eit}{\end{itemize}}
\newcommand{\bd}{\begin{description}}
\newcommand{\ed}{\end{description}}

\newcommand{\bc}{\begin{center}}
\newcommand{\ec}{\end{center}}
\renewcommand{\Ref}[1]{(\ref{#1})}

\newcommand{\C}{{\mathbb C}}
\newcommand{\N}{{\mathbb N}}
\newcommand{\R}{{\mathbb R}}
\newcommand{\Z}{{\mathbb Z}}

\newcommand{\cD}{{\mathcal D}}

\newcommand{\SU}{\mathrm{SU}}

\newcommand{\SL}{\mathrm{SL}}
\newcommand{\SO}{\mathrm{SO}}

\newcommand{\su}{{\mathfrak{su}}}

\newcommand{\be}{\begin{equation}}
\newcommand{\ee}{\end{equation}}
\newcommand{\bea}{\begin{eqnarray}}
\newcommand{\eea}{\end{eqnarray}}
\newcommand{\bs}{\begin{subequations}}
\newcommand{\es}{\end{subequations}}
\newcommand{\nn}{\nonumber}

\newcommand{\tr}{{\rm Tr}}
\newcommand{\f}{\frac}

\def\p{\partial}
\newcommand{\Id}{\mathds{1}}

\newcommand{\scr}{\scriptscriptstyle\rm}

\newcommand{\ra}{\rangle}
\newcommand{\la}{\langle}
\newcommand{\bra}[1]{\langle {#1}|}
\newcommand{\ket}[1]{|{#1}\rangle}
\newcommand{\mean}[1]{\langle{#1}\rangle}

\newcommand{\mat}[4]{\left( \begin{array}{cc}{#1}&{#2}\\{#3}&{#4}\end{array} \right)}

\renewcommand{\d}{\delta}  \newcommand{\eps}{\epsilon}  \newcommand{\z}{\zeta}
       
\let\m=\mu     \newcommand{\s}{\sigma}  \renewcommand{\t}{\tau}     \let\om=\omega
\let\G=\Gamma \let\D=\Delta     \let\Om=\Omega

\newcommand{\norm}[1]{|\!|#1|\!|}

\newcommand{\Wthree}[6]{\left(\begin{array}{ccc} #1 & #2 & #3 \\ #4 & #5 & #6 \end{array}\right)}
\newcommand{\Wfour}[9]{\left(\begin{array}{cccc} #1 & #2 & #3 & #4 \\ #5 & #6 & #7 & #8 \end{array}\right)^{(#9)}}

\newcommand{\wh}{\widehat}
\newcommand{\sL}{{\scr L}}
\newcommand{\sR}{{\scr R}}

\newcommand{\exiop}{\widehat{e^{i\xi}}}

\usepackage[normalem]{ulem}

\def\hS{\widehat{S}}
\def\hT{\widehat{T}}

\begin{document}

\title{\bf Twisted Geometries Coherent States \\ for Loop Quantum Gravity}

\author{\Large{Andrea Calcinari$^{a,b}$, Laurent Freidel$^c$, Etera Livine$^d$ and Simone Speziale$^a$}
\smallskip \\
\small{$a$ Aix Marseille Univ., Univ. de Toulon, CNRS, CPT, Marseille, France}\\
\small{$b$ DIFA, Alma Mater Studiorum, Universit\`a di Bologna, Italy}\\
\small{$c$ Perimeter Institute for Theoretical Physics, 31 Caroline St. N, N2L 2Y5, Waterloo ON, Canada}\\
\small{$d$ Universit\'e de Lyon, ENS de Lyon, Universit\'e Claude Bernard, CNRS, LP ENSL, 69007 Lyon, France}
}
\date{\today}

\maketitle

\begin{abstract}
\noindent We introduce a new family of coherent states for loop quantum gravity, inspired by the twisted geometry parametrization. We compute their peakedness properties and compare them with the heat-kernel coherent states. They show similar features for the area and the holonomy operators, but improved peakedness in the direction of the flux. At the gauge-invariant level, the new family is built from tensor products of coherent intertwiners. To study the peakedness of the holonomy operator, we introduce a new shift operator based on the harmonic oscillator representation associated with the twisted geometry parametrization. The new shift operator captures the components of the holonomy relevant to disentangle its action into a simple positive shift of the spins.
\end{abstract}

\tableofcontents

\section{Introduction}

The quantum geometry described by loop quantum gravity (LQG) has non-commutative properties \cite{RovelliSmolin94,Ashtekar:1998ak,Thiemann:2000bw}: these include the incompatibility of holonomies and fluxes as conjugate pairs,  and also of different flux components. As a result, discussing the semiclassical limit of loop quantum gravity requires the use of coherent states to minimize the effects of such non-commutativity. After early constructions \cite{Ashtekar:1992tm,Ashtekar:1994nx}, a complete framework was provided by Thiemann's 
heat-kernel (or complexifier) coherent states \cite{Thiemann:2000bw,Hall,Thiemann:2000ca,Thiemann:2002vj}.
These states have a number of key properties, and have played a major role in loop quantum gravity. 
On a fixed graph, the heat-kernel coherent states give well-peaked distributions for the holonomies and commuting flux components \cite{Thiemann:2000ca}. They don't give on the other hand well-peaked distributions for
the direction of the fluxes. They are built, after all, in the purpose of minimizing other non-commuting operators rather than the different flux components. As a consequence, 
their gauge-invariant projection does not contain the coherent intertwiners as introduced in \cite{LS}, that use minimal direction uncertainty to provide each intertwiner with the geometric interpretation of a polyhedron \cite{EteraHoloQT,EteraFreidelUN,IoPoly}.
Only the subset of heat-kernel coherent states with large area labels (namely large values of the norm of the fluxes) pick out precise flux directions and reproduce the coherent intertwiners at the gauge-invariant level \cite{Bianchi:2009ky}. The lack of precise direction of the heat-kernel coherent states is not a problem per se, just a fact resulting from choosing a certain coherent state family with other peakedness properties. Furthermore, it 
 can be argued that only the states with large area labels are truly semiclassical anyways, and for those the directions are indeed well-peaked. 
Nonetheless, a number of applications like spin foam dynamics, group field theory and quantum reduced models, indicate the usefulness to have a clear flux direction also at finite area. Specifically, it is useful to have access to a resolution of the identity in which every element gives a clear direction, without spoiling the peakedness in the holonomy components. That it is possible to achieve peakedness in both holonomies and flux-directions is not obvious a priori, and we show here that this is the case.

We introduce a new family of coherent states that has these two  properties: 
First, it is well-peaked on the flux directions for all values of the area label and, second,
its gauge-invariant projection contains tensor products of coherent intertwiners for all values of the area label.
This result shows that coherent intertwiners are a useful basis for the full Hilbert space of LQG, and not just for a single node. 
 The new family  is based on the twisted geometry parametrization, and indeed most of the work reported here was done shortly after \cite{twigeo}. 
 We were motivated to complete the study by the recent developments in effective dynamics for loop quantum gravity derived from the heat-kernel coherent states   \cite{Dapor:2017rwv,Alesci:2018loi,Li:2018opr,Liegener:2019dzj} 
(for previous work see \cite{Bojowald:2005cw,Yang:2009fp,Qin:2012xh}). 
We think in fact that having an alternative family of coherent states available should be of help in assessing the model-independence of such scenarios, and to provide alternative geometric description of the data based on twisted geometries. This alternative geometric description is particularly useful at the gauge-invariant level, where it allows a complete characterization of the reduced variables as a collection of polyhedra with a notion of intrinsic and extrinsic curvature.

To make our paper more self-contained, we start in Section 2 with a brief presentation of the peakedness properties of the heat-kernel coherent states, computing expectation values and relative uncertainties of the fluxes and holonomies. These results are well-known to experts, but not always easily accessible in the literature, which tends to refer to probability distributions only. The evaluations are somewhat cumbersome for the holonomy operator. To improve the situation, we introduce a new shift operator, that disentangles the holonomy action into a simple positive shift of the spins and magnetic numbers. This operator can be constructed by exploiting the harmonic oscillator representation of the holonomy-flux algebra, defined by the spinorial and twistorial parametrization \cite{twigeo2,Livine:2011gp,IoWolfgang,Bianchi:2016hmk}. It captures the component of the holonomy that cannot be reconstructed having access to both left and right-invariant vector fields, and we believe it can have useful applications beyond the present ones.

In Section 3 we introduce the new family of twisted geometries coherent states, which we construct by taking tensor products of $\SU(2)$ coherent states for both left and right-invariant vector fields, with a spin weight similar to the heat-kernel coherent states. 
We compute the expectation values and relative uncertainties, and compare them with the heat-kernel coherent states. 
The peakedness in the area and in the shift operator are qualitatively the same, while a stronger peakedness in the flux direction is achieved.
The presence of SU(2) coherent states guarantees a clear identification of the directions, and an exact relation to coherent intertwiners at the gauge-invariant level. 
It is not possible on the other hand to also obtain a strict minimization of the uncertainty between the different flux components, because this property of the SU(2) coherent states is spoiled by the sum over the spins needed in the full Hilbert space of LQG.

The price to pay for these new properties is that the states we write are not eigenstates of an annihilation operator, at least not one that we were able to identify. As a consequence, it is not known if and which Heisenberg relation they saturate.
Hence, they are coherent states in a weaker sense: they provide a resolution of the identity and have good peakedness properties on a point in phase space, but they are not eigenstates of an annihilation operator. 
They could be referred to as generalized coherent states, following the notation and classification of \cite{GazeauBook}.
We also introduce a second family of twisted geometries coherent states, simpler-looking but less symmetric, and discuss its pros and cons, as well as the general idea behind our construction. Finally, we also compare the states with the spinorial coherent states defined in \cite{EteraFreidelUN,EteraSpinor,EteraTamboSpinor,Dupuis:2010iq,Bonzom:2012bn,Alesci:2015yqa} and used in spin foam amplitudes in \cite{EteraHoloEucl,Hnybida:2014mwa,Banburski:2014cwa,IoHolo,Hnybida:2015ioa} and for entanglement calculations on spin networks \cite{Bianchi:2018fmq}. 

The Appendix contains a complete formularium of conventions for twisted geometries, the harmonic oscillator representation of the holonomy-flux algebra, and coherent intertwiners.

\section{Background: Heat-kernel coherent states}

We will be discussing coherent states for loop quantum gravity on a fixed, oriented graph. 
The building block is the Hilbert space $L_2[\SU(2)]$ of square-integrable functions on SU(2) with respect to the Haar measure.
It is associated to each oriented link of the graph, and it carries a representation of the holonomy-flux algebra.
We define the flux operators as right-invariant vector fields $\hat R$ acting on a group element $g$ and associated with the source node of the link.
They are related by the adjoint transformation to the left-invariant vector fields $\hat{L}$, 
associated with the target node of the link. 
Two convenient basis are the holonomy basis $\ket g$ of eigenvectors of the holonomy operator, and the 
momentum basis $\ket{j,m,n}$
which diagonalizes the z-components $\hat R_z$ and $\hat L_z$, and the common Casimir $\hat{L}^2$ to both left and right-invariant operators.
 The overlap between the two basis is given by the Wigner matrices, and we take conventions
\be
\bra{\bar g}j,m,n\ra = \sqrt{d_j} D^{(j)}_{mn}(g).
\ee
A full list of conventions and useful results is summarized in Appendix~\ref{AppA}. Gauge-invariant states on the graph are obtained by group averaging, and we write the resulting spin networks schematically as 
\be
\bra{\bar g_l}\G,j_l,i_n\ra = \tr_\G  \otimes_l \sqrt{d_j} D^{(j)}(g_l) \otimes_n i_n,
\qquad \bra{\G,j_l,i_n}\G,j'_l,i'_n\ra = \d_{j_l,j_l'}\d_{i_n,i'_n}.
\ee
Here $i_n$ is used both as the intertwiner tensor state and intertwiner label in a recoupling basis, an abuse of notation common in the literature,
and $\ket{\G,j_l,i_n}$ are normalized spin networks. The trace $\tr_\G$ is over the magnetic index structure dictated by the graph connectivity, and assures gauge invariance.
symbol $\tr_\G$
An arbitrary gauge-invariant state can be decomposed in the spin-network basis as
\be
\Psi(g_l) = \sum_{j_l,i_n} \Psi_{j_l,i_n} \la \bar g_l \ket{\G,j_l,i_n}.
\ee

\subsection{Parametrizations of $T^*\SU(2)$}
The semiclassical limit of $L_2[\SU(2)]$ with its holonomy-flux algebra is the manifold $T^*\SU(2)$ with its canonical SU(2)-invariant symplectic structure. Therefore coherent states for $L_2[\SU(2)]$ are labelled by points in $T^*\SU(2)\simeq \SU(2)\times\su(2)$, namely pairs $(g,L)$ of group and algebra elements, the latter identified by convention with the left-invariant vector fields acting on $\SU(2)$ as a manifold. 
The usual geometric interpretation of these labels for coherent states is in terms of distributional geometric data, corresponding to the holonomy of the Ashtekar-Barbero connection along the link, and fluxes, namely integrals of the densitized triad vector, over chosen surfaces dual to the links
\cite{ThiemannBook}. There are two convenient alternative parametrizations of this symplectic manifold. The first exploits the isomorphism $T^*\SU(2) \simeq \SL(2,\C)$ given by the polar decomposition
\be
H:=e^{iL}g \in \SL(2,\C).
\ee
The second exploits the adjoint relation between left and right-invariant vector fields $R=-gLg^{-1}$ to trade holonomies and fluxes for two unit vectors, one norm, and an angle: 
\begin{subequations}\label{twigeo}\begin{align}
& (g,L) \mapsto (A,\xi, \z_{\sL}, \z_{\sR}), \qquad A\geq 0, \quad \xi\in[-2\pi,2\pi), \quad \z_\sL\in\C P^1,  \quad \z_\sR\in\C P^1, 
\\\label{tgdomain}
& g=n_{\sL}e^{\xi\t_3}n_{\sR}^{-1}, \qquad L = An_\sL \t_3 n_\sL^{-1}, \qquad R= -An_\sR \t_3 n_\sR^{-1}. 
\end{align}\end{subequations}
Here $n_{\sL,\sR}=n(\z_{\sL,\sR}):\C P^1\mapsto\SU(2)$ is a section in the Hopf fibration, and $\t_3=-i\s_3/2$. The \emph{twist angle} $\xi$ should not be confused with the class angle of the more common polar parametrization of SU(2), as can be immediately seen taking the trace of $g$. In particular, its range is twice as big, see Appendix~\ref{AppA} for details.
This parametrization has two useful properties. First, it identifies the twist angle as a choice of canonically conjugated variable to the spin.
Second, it simplifies the solution to the closure constraints defining the gauge-invariant phase space on a graph. The result is a geometric interpretation of the gauge-invariant variables in terms of twisted geometries. This is a piecewise flat geometric interpretation of the labels, alternative to the distributional one, and based on a specific choice of dual surfaces. 

These are two parametrizations of the same phase space. For that, the domain of the variables is crucial. It is given by the whole of $\SL(2,\C)$ in the first case, and by \Ref{tgdomain} in the second.
The first parametrization is at the heart of the heat-kernel coherent states, which are obtained via holomorphic quantization associated with the $\SL(2,\C)$ complex structure. The second parametrization is is at the heart o the new family of coherent states proposed in this paper.
The relation between the two parametrizations is given by
\be\label{Htg}
H = e^{iL}g = n_\sL e^{\om\t_3} n^{-1}_\sR,
\ee
where we introduced
\be\label{defom}
\om:=\xi+iA.
\ee

\subsection{Heat-kernel coherent states}

The heat-kernel, or complexifer coherent states are given by \cite{Thiemann:2000bw} (see also \cite{Hall,Thiemann:2000ca,Thiemann:2002vj,Bahr:2007xn})
\be\label{HKCS}
\Psi^t_{H}(g) = \bra{\bar g}H,t\ra 
= \sum_j d_j e^{-\f t2j(j+1)} \chi^{(j)}(Hg).
\ee
Here $H\in\SL(2,\C)$ is the classical label, and $\chi^{(j)}$ the SU(2) character.\footnote{Thiemann's original definition uses a $g^{-1}$ argument, namely
\[
\sum_j d_j e^{-\f t2j(j+1)} \chi^{(j)}(Hg^{-1}).
\]
It is related to \Ref{HKCS} by a redefinition of $H$, motivated by the different conventions used for the parametrization of $T^*\SU(2)$.
}
The classicality parameter $t\in\R^+$ controls the relative spread between holonomies and fluxes, and plays an important role in adapting the family of coherent states to the semi-classical applications. As we review below, small values of $t$ are required for good peakedness properties.
It is also an imprint of the kinematic nature of the states: one may expect that the dynamics fixes the spread of the physical coherent states.
See the cited literature for a more complete discussion of this parameter.
The norm of the states is 
\be
\norm{\Psi_H^t}^2 = \sum_j d_j e^{-tj(j+1)} \chi^{(j)}\left( H^\dagger H\right)
= \sum_j d_j e^{-t j(j+1)} \f{{\sinh d_j A}}{\sinh A}. 
\ee
In the last equality we used \Ref{Htg}, to highlight that the norm depends only on the boost rapidity $A$, which geometrically has the interpretation of an area. We can also write them conveniently as kets, using the angular momentum or electric basis
\be
\ket{H,t}= \sum_{jmn}d_j^{\f 12} e^{-\f t2j(j+1)} D^{(j)}_{nm}(H)\ket{j,m,n}.
\ee

These coherent states have a number of important properties that make them useful to LQG. Three of their properties are relevant to the present paper.
$(i)$ They are eigenstates of a matrix of annihilation operators, given by $\hat H =e^{i\hat L}\hat g$ \cite{Thiemann:2000bw};  and thus they minimize the uncertainty relation associated with the corresponding three operators. $(ii)$ They provide an over-complete basis, with resolution of the identity given by an integral in phase space,
\be
\Id=\int_{T^*\SU(2)}d\m(H)\ket{H,t}\bra{H,t},
\ee
with
\be
\int_{T^*\SU(2)}d\m(H):= \f{e^{-\f t4}}{(\pi t)^{\f32}} \int_{\R^3} d^3L\f{\sinh |L|}{|L|}e^{-\f{|L|^2}t} \int_{\SU(2)}d g.
\ee
$(iii)$ They also peak the holonomy and fluxes on the classical values $H=g e^{iL}$, with vanishing relative uncertainties for $t\rightarrow 0$ and $A\gg t$.
This was shown in \cite{Thiemann:2000ca} through an extensive analysis
of the distributions $\f{|\Psi_H^t(g)|^2}{\norm{\psi_H^t}^2}$ and $\f{|\Psi_H^t(j,m,n)|^2}{\norm{\psi_H^t}^2}$, identifying their peaks and spreads. 

From this building block we can define coherent states for the gauge-invariant Hilbert space 
${\cal H}_0:=L_2[\SU(2)^L/\SU(2)^N]$ associated with a graph $\G$ with $N$ nodes and $L$ oriented links. This is obtained by group averaging, and can be compactly written as
\begin{align}
{\cal H}_0 \ni \ket{\G,H_l,t} &
= \sum_{j_l,i_n} \bigg(\prod_l d^{\f12}_{j_l} e^{-\f t2{j_l(j_l+1)}}\bigg) \tr_\G \left[ \otimes_l D^{(j_l)}(H_l) \otimes_n i_n\right]\ket{\G,j_l,i_n}.
\end{align}
Notice the entangled structure of the trace: this is a complexity that hinders certain explicit calculations and which is simplified by our new family of coherent states.

\subsection{Large area limit}

Replacing $H$ with the twisted geometry parametrization through \Ref{Htg},
it is easy to show that
\be\label{HKlimit}
\lim_{A\rightarrow\infty} \Psi^t_H(g) 
= e^{\f 1{2t}( A -\f t2)^2}\sum_j d_j e^{-\f t2(\f{d_j}2-\f At)^2-i\xi j} D^{(j)}_{jj}(n_\sR^{-1}g  n_\sL).
\ee
This follows from $D^{(j)_{mn}}(e^{\om\t_3})=e^{-i\om m}\d_{mn}$, and
\be\label{Gfactor}
-\f t2{j(j+1)}+(A-i\xi_l) j = \f 1{2t}\left( A -\f t2\right)^2- \f t2\left(\f{d_j}2-\f At\right)^2-i\xi j.
\ee
Equivalently in the momentum basis,
\be
\lim_{A\rightarrow\infty}\ket{H,t} 
= e^{\f 1{2t}( A -\f t2)^2}\sum_j d_j^{\f 12} e^{-\f t2(\f{d_j}2-\f At)^2-i\xi j} D^{(j)}_{nj}(n_\sL) D^{(j)}_{jm}(n_\sR^{-1})\ket{j,m,n}.
\ee
The large area limit projects the magnetic indices on their highest weights, and the heat-kernel states factorize in tensor products of SU(2) coherent states. Furthermore, the conjugate pair $(A,\xi)$ is peaked with a simple Gaussian structure, which reproduces the ansatz used in the spin foam graviton propagator calculations \cite{Bianchi:2006uf,IoASL}.\footnote{In that context, the parameter $t$ can be fixed with dynamical considerations, see e.g. \cite{Livine:2007mr,Dittrich:2007wm,Dupuis:2011um}.}
This limiting behaviour was pointed out in \cite{Bianchi:2009ky} and used to make a first contact between the heat-kernel coherent states and the coherent intertwiners. 
In fact, it follows  from \Ref{HKlimit} that at the gauge-invariant level, 
\begin{align}
\lim_{A\rightarrow\infty} \ket{\G,H_l,t} =
\sum_{j_l,i_n} \prod_l d^{\f12}_{j_l} e^{-\f t2{j_l(j_l+1)}+(A_l-i\xi_l) j_l} 
\prod_n c_{i_n}(n_l) \ket{\G,j_l,i_n}
\end{align}
where $c_i(n_l)$ are the coefficients of the coherent intertwiners \cite{LS}, depending on the relevant coherent state labels $n_\sL$ or $n_\sR$ associated with the links $l$ touching the node $n$, see Appendix~\ref{AppCI} for conventions.
Since coherent intertwiners appear only in the large $A$ limit, the resolution of the identity of the heat-kernel coherent states includes gauge-invariant states which do not have the nice geometric interpretation of fuzzy polyhedra. The lack of minimal spread in the direction for finite $A$ will be shown more explicitly below.

To construct states always peaked on the direction like the SU(2) coherent states, one may first try with a simple Gaussian like in the right-hand side of \Ref{HKlimit} as the definition of a new family of coherent states, namely
\be\label{PsiG?}
\Psi{}^t_{\scr G}(g)\stackrel{?}{:=}\sum_j d_j e^{-\f t2(j-A)^2-i\xi j} D_{jj}(n_\sR^{-1}g  n_\sL),
\ee
to be valid for any $A$. 
This however would not work, as it is not possible to find a resolution of the identity with these states. The difficulty comes from the fact that the phase space is spanned by $A\in\R^+$, and there is no temperate distribution $\m_t(A)$ that would satisfy
\be
\int_{0}^{+\infty}d A\,\mu_{t}[A]\,e^{-d_{j}A}\stackrel{?}{=}e^{\f14 td_{j}^2}.
\ee
This can be understood observing that the LHS is a Laplace transform, whose inverse is given by the Mellin transform with integration over vertical lines in the complex plane, and it is well-known that a positive Gaussian weight $e^{\f14 td_{j}^2}$, which grows to infinity as $j\rightarrow +\infty$, does not admit any inverse Laplace transform (see e.g. \cite{debnath2014integral}, p.164).
At an intuitive level, one sees that the left hand side integral decreases in $d_{j}$ assuming that the integration measure $\mu_{t}[A]$ is positive, while the right hand side positive Gaussian increases and diverges in $d_{j}$ as soon as $t>0$. This can be made mathematically precise for an absolutely convergent Laplace transform.

A possible solution is to introduce a stronger damping factor,
\be\label{PsiG}
\Psi{}^t_{\scr G}(g):=\sum_j d_j \f{ e^{-\f t2(j-A)^2-i\xi j} }{\sqrt{1+{\rm Erf}(j\sqrt{t})}} D_{jj}(n_\sR^{-1}g  n_\sL).
\ee
This option would provide a resolution of the identity, with measure 
\be\label{Gaussianmeasure}
\Id = 2 \sqrt{\f t\pi} \int_{\R^+} {dA} \int_{-2\pi}^{2\pi} \f{d\xi}{4\pi} \int_{\C P^1} d\m(\z_{\sL}) \int_{\C P^1} d\m(\z_{\sR}).
\ee
The main result of our paper is to show that there exists a more natural choice, with better peakedness properties than those entailed by the 
ad hoc Erf function in \Ref{PsiG}.

\subsection{Expectation values and peakedness }
In this Section we discuss in more details the expectation values of the heat-kernel coherent states, briefly summarized by property $(iii)$ above.
It will be useful
for comparison to our states below. We look separately at the expectation values and relative uncertainties of the spin,  of the fluxes, and of the holonomy.

\subsubsection*{Spin }
\begin{figure}[H]
\centering
\includegraphics[width=7.8cm]{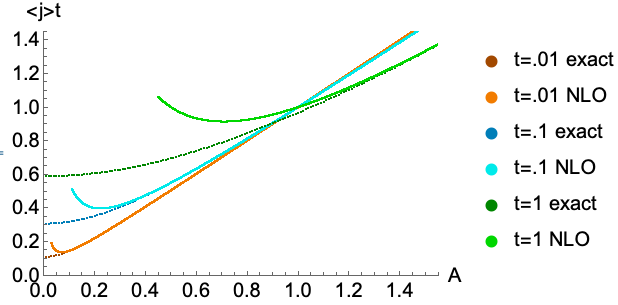}\hspace{.4cm}\includegraphics[width=8cm]{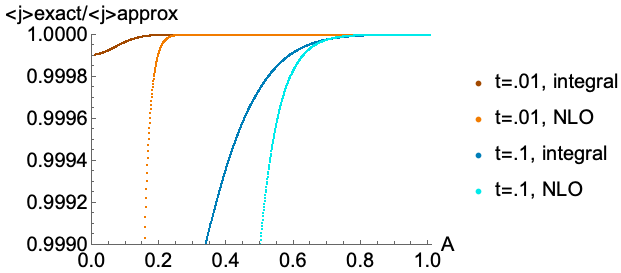}\vspace{.5cm}
\includegraphics[width=8cm]{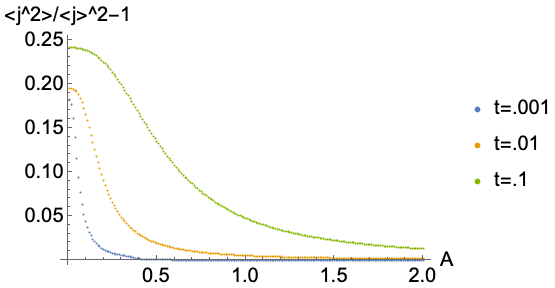}
\caption{\small{\emph{Numerical studies of the spin peakedness with heat-kernel coherent states}. Top, left panel: \emph{Numerical evaluations of \Ref{EVjT} for different values of $t$, with the NLO approximation.} Top, right panel: \emph{Further studies of the expectation value, comparing the integral and NLO approximations.}
Bottom panel: \emph{The exact relative uncertainty, vanishing for $A\gg t$.}}\label{FigEVjT}
}
\end{figure}

For the spin distribution,
\cite{Thiemann:2000ca} gives the estimate
\be\label{electricdistr}
\f{|\Psi_H^t(j,m,n)|^2}{\norm{\psi_H^t}^2} \lesssim \f{t^{\f 32}}{\sqrt \pi} \f{j+1/2}{1-K_t} \exp\left(-t\Big(\f{d_j}2-\f At\Big)^2  
\right),
\ee
declared valid for all magnetic labels, and large spins. Here $K_t$ is an undetermined positive constant decaying exponentially to zero as $t\rightarrow 0$.
We see a peak at $j=A/t-1/2$, with spread $1/t$. Since the location of the peak grows with $1/t$ as well, the relative uncertainty scales as $\sqrt{t}$ and vanishes for $t\rightarrow 0$. This estimate can be compared with the exact 
expectation value
\be\label{EVjT}
\mean{\hat\jmath}:=\f{\bra{H,t} j\ket{H,t}}{\norm{\psi_H^t}^2} = \f{\sum_j j d_j e^{-t j(j+1)}{\sinh d_j A}}{\sum_j d_j e^{-t j(j+1)} {\sinh d_j A} },
\ee
which is plotted in Fig.~\ref{FigEVjT}. It confirms the value of the peak for large $A$, but with departures at small $A$.
We conclude that the estimate \Ref{electricdistr} should be used only for large $A$, consistently with large spins.

An analytic expression of \Ref{EVjT}  can be obtained if we approximate the sums with an integral in $\R^+$. This approximation can be justified for small $t$, using Poisson's resummation formula. The integral can be performed exactly in terms of Erf functions, and an expansion for large $A$ gives 
\be\label{EVjTLO}
\mean{\hat\jmath} \simeq \f At-\f 12 +\f{1}{2A}.
\ee
The leading order of this approximation is consistent with \Ref{electricdistr}, and we have obtained a next-to-leading order (NLO) correction. 
Using the same approximations for the variance, we obtain 
\be\label{RUj}
\f{\mean{\hat\jmath^2}}{\mean{\hat\jmath}^2}-1\simeq \f t{2A^2}.
\ee
The relative uncertainty vanishes for $A\to\infty$. Numerical tests show that the NLO formula is very accurate 
for $A\gg t$, 
and provide a profile for the relative uncertainty at finite $A$, see Fig.~\ref{FigEVjT}.

There are two useful remarks to make at this point. First, only the coherent states with $A\gg t$ have good semiclassical properties, namely the distributions are peaked on the classical phase space point, with vanishing relative uncertainties. As explained in \cite{Thiemann:2000ca}, this fact is due to the non-linear structure of the group manifold. Second, the expectation value of the spin in this limit is $A$ rescaled by $1/t$. For this reason the label of the semiclassical states is often renamed $p:=A/t$. 

\subsubsection*{Fluxes}
Including the magnetic numbers, \cite{Thiemann:2000ca} gives the estimate
\be\label{electricdistr1}
\f{|\Psi_H^t(j,m,n)|^2}{\norm{\psi_H^t}^2} \lesssim\f 1A \f{t^{\f 32}}{4\sqrt \pi} \f{1}{1-K_t} \exp\left(-t\Big(\f{d_j}2-\f At\Big)^2  
{-\f j2\f{(m/j-R_z/A)^2}{1-(R_z/A)^2}}-\f j2\f{(n/j-L_z/A)^2}{1-(L_z/A)^2}
\right),
\ee
 declared this time to be valid only for large $A$. This is the expression given for $L_z/A<1$ and $R_z/A<1$, and there is a similar one otherwise. As a  distribution, it implies that the $z$-components of both left-invariant and right-invariant fluxes are well peaked, at least for $t\rightarrow 0$ and $A\to\infty$, as indeed confirmed by the numerics. 
There is however not much information about the direction, and indeed the $x$ and $y$ components of the fluxes are completely spread. This can be seen by computing expectation values. For the left-invariant  flux for instance, we have
\begin{align}\label{HKLev}
& \bra{H,t}\hat{\vec L}\ket{H,t} = \sum_j d_j e^{-t C_j} \sum_m e^{2A m} \bra{j,m}n_\sL^\dagger {\vec L} n_\sL\ket{j,m}, \\
& \bra{H,t}\hat{L_i^2}\ket{H,t} = \sum_j d_j e^{-t C_j} \sum_m e^{2A m} \bra{j,m}n_\sL^\dagger {L^2_i} n_\sL\ket{j,m}. \label{HKLqev}
\end{align}
The lack of sharp direction properties is made manifest by the presence of matrix elements of $\vec L$ with kets with non-minimal spread, namely those with $m\neq\pm j$. 

The situation improves  in the large $A$ limit, when the magnetic sum above is dominated by the highest term.
We can then use the well-known properties of SU(2) coherent states
\be\label{nLn}
D_{\pm j\pm j}(n^\dagger \vec L n) = \pm j\vec n,\qquad D_{\pm j\pm j}(n^\dagger L_i^2 n) = j^2n_i^2+\f j2(1-n_{i}^2),
\ee
with $\vec n$ is the unit vector pointing in the direction $\z$ on the sphere.
The normalized expectation value gives 
\be
\mean{\hat{\vec L}}:= \f{\bra{H,t}\hat{\vec L}\ket{H,t} }{\norm{\Psi_H^t}^2} \stackrel{A\rightarrow \infty}{\longrightarrow} 
\f{\sum_j j d_j e^{-t C_j} e^{2 jA}}{\sum_j d_j e^{-t C_j} e^{2 jA} } \vec n_\sL \simeq \Big(\f At-\f 12 +\f{1}{2A}\Big) \vec n_\sL, 
\ee
using the integral approximation valid at small $t$. Proceeding similarly for the variance,
we find
\be
\D^2 \hat L_i=\f{\mean{\widehat{L_i^2}}}{\mean{\hat L_i}^2}-1 \simeq \f{1-n_i^2}{n_i^2}\f t{2A}, 
\ee
vanishing in the large area limit. This behaviour is consistent with \Ref{electricdistr1}, and further shows that the mean of each component of the flux is well-peaked on the classical label at large area, and not just the $z$ component.
It can also be understood simply as a direct consequence of the projection of the heat-kernel states on SU(2) coherent  states discussed earlier in \Ref{HKlimit}. 

\subsubsection*{Holonomy} 

The peakedness of the holonomy operator for $t\rightarrow 0$ can be  seen from the definition \Ref{HKCS}: in this limit the state approaches a delta distribution. A more detailed analysis \cite{Thiemann:2000ca}, showing in particular that the presence of a boost in $H$ does not spoil the 
behaviour of the approximate delta function, 
gives again a Gaussian distribution, with spread $\sqrt{t}$.
We refrain from reporting here these results, and rather point out that a 
simpler and more explicit estimate of the peakedness properties on holonomy components can be obtained looking at the shift operator suggested by twisted geometries, namely the exponential of the twist angle $\xi$. This operator can be defined with the following choice of ordering,
\be\label{exiop}
\exiop := (n_\sR^0)^{-1} (a_\sR^{0\dagger})^2 (n_\sL^0)^{-1} (a_\sL^0)^2,
\qquad 
\exiop \ket{j,m,n} = \f{(j+m+1)^{1/2}}{(j+m+2)^{1/2}} \f{(j+n+1)^{1/2}}{(j+n+2)^{1/2}} \ket{j+1,m+1,n+1}.
\ee
See Appendix~\ref{AppShift} for more details.
From this action we find 
\begin{align}
\bra{H,t} \wh{e^{i\xi}} \ket{H,t}&=\sum_j d_j e^{-t(j+1)^2} \left( \f{2j+3}{2j+1}\right)^{\tfrac12}
\sum_{m,n} \left( \f{j+m+1}{j+m+2}\f{j+n+1}{j+n+2}\right)^{\tfrac12}
\\\nn &\qquad \times 
\sum_{r=-j}^j e^{-i\om r} D^{(j)}_{nr}(n_{\scr L})D^{(j)}_{rm}(n_{\scr R}^\dagger)
\sum_{p=-j-1}^{j+1} e^{i\bar\om p} D^{(j+1)}_{n+1,p}(\bar n_{\sL})D^{(j+1)}_{m+1,p}(n_{\sR}).
\end{align}
The functions
\be
\Xi^j_{nm}(\om,n_1,n_2):=\sum_{r=-j}^j e^{-i\om r} D^{(j)}_{nr}(n_1)D^{(j)}_{rm}(n_2)
\ee
are known explicitly from the Wigner matrices. 
These and their counterparts for $\wh{e^{2i\xi}}$ can be used to study the expectation value and peakedness.

\begin{figure}[H]
\centering
\includegraphics[width=8cm]{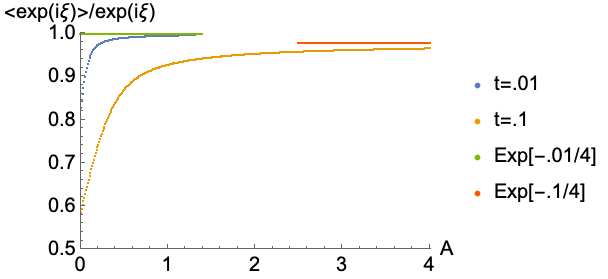}\hspace{.1cm} \includegraphics[width=8cm]{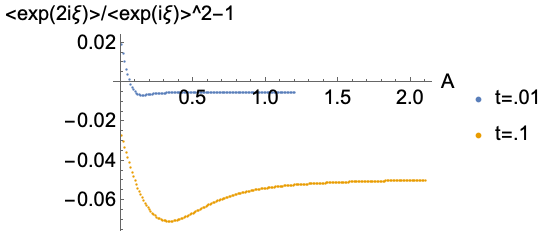}
\caption{\small{\emph{Numerical studies of the twist angle expectation value with heat-kernel coherent states.
The normalized expectation value \Ref{EXxiT} tends to $e^{i\xi}e^{-t/4}$ for $A\gg t$} (left panel), \emph{and the relative uncertainty to $-t/2$.} (right panel). }
\label{FigEVxiT}
}
\end{figure}

To gain more insight, we consider the simplest example $n_\sL=\Id=n_\sR$, namely $H=e^{\om\t_3}$. In this case, 
$\Xi^j_{nm}(\om,\Id,\Id)=e^{-i\om m}\d_{mn}$ and
\begin{align}\label{EVxiT1}
\bra{e^{\om\t_3},t} \wh{e^{i\xi}} \ket{e^{\om\t_3},t}&=e^{i\xi}\sum_j d_j e^{-t(j+1)^2} \left( \f{2j+3}{2j+1}\right)^{\tfrac12}
e^{A (1-2j)}  \sum_{k=0}^{2j} \f{k+1}{k+2} e^{2Ak}.
\end{align}
The classical value $e^{i\xi}$ now clearly factorizes in the average. 
A similar calculation for the square gives
\begin{align}\label{EVxiT2}
\bra{e^{\om\t_3},t} \wh{e^{2i\xi}} \ket{e^{\om\t_3},t}&=e^{2i\xi}\sum_j (2j+1)^{\f12}(2j+5)^{\f12} e^{-t(j^2+3j+3)} 
e^{A (2-2j)}  \sum_{k=0}^{2j} \f{(k+1)(k+3)}{(k+2)(k+4)} e^{2Ak}.
\end{align}
Taking only the dominating term in the large $A$ limit, we find
\be\label{EXxiT}
\mean{ \wh{e^{i\xi}} } := \f{\bra{e^{\om\t_3},t} \wh{e^{i\xi}} \ket{e^{\om\t_3},t}}{\norm{\psi_H^t}^2} \sim e^{i\xi} e^{-\f t4}+O(A^{-1}),
\qquad \mean{ \wh{e^{2i\xi}} } := \f{\bra{e^{\om\t_3},t} \wh{e^{2i\xi}} \ket{e^{\om\t_3},t}}{\norm{\psi_H^t}^2} \simeq e^{2i\xi} e^{-t}+O(A^{-1}).
\ee
These estimates are well supported by numerical studies, reported in Fig.~\ref{FigEVxiT}.
The dependence on $\xi$ is consistent with the semiclassical behaviour expected from the coherent states, but we notice a mismatch by an exponential of $t$, inducing an imaginary shift of $\xi$. This situation  is  familiar from the particle in a circle \cite{Kowalski}, 
for which $\mean{\wh{e^{in \xi}} } = \exp(-t/4 n^2) \exp in\xi$.
As a consequence, the relative uncertainty decreases for large $A$, but never really vanishes, reaching a non-zero minimal value for any non-zero value of $t$,
\be
\D^2 \widehat{e^{i\xi}}:=\f{\mean{\widehat{e^{2i\xi}}}}{\mean{\widehat{e^{i\xi}}}^2}-1= e^{-\f t2}-1\simeq -\f t2.
\ee

Another case where it is possible to obtain  compact expressions is for $A\rightarrow\infty$. This limit
projects the magnetic indices on their highest values, giving 
\begin{align}
\lim_{A\rightarrow\infty}\bra{H,t} \wh{e^{i\xi}} \ket{H,t}&= 
e^{i\xi} \sum_j d_j^{\f12} (2j+3)^{\f12} e^{-t(j+1)^2}  e^{d_j A} 
\sum_{m,n} \left( \f{j+m+1}{j+m+2}\f{j+n+1}{j+n+2}\right)^{\tfrac12}
\\\nn &\qquad \times D^{(j)}_{nj}(n_{\scr L})D^{(j)}_{jm}(n_{\scr R}^\dagger) D^{(j+1)}_{n+1,j+1}(\bar n_{\sL})D^{(j+1)}_{m+1,j+1}(n_{\sR}),
\end{align}
with the desired semiclassical value clearly factorized. The Wigner matrices can be evaluated using \Ref{HopfSection} and \Ref{WignerHighest} in Appendix~\ref{AppA}, and we find
\begin{align}
& \sum_m \left( \f{j+m+1}{j+m+2} \right)^{\tfrac12} D^{(j)}_{jm}(n_{\sR}^\dagger)  D^{(j+1)}_{m+1,j+1}(n_{\sR})
= \f{ {\cal D}_j(|\z_\sR|^2) }{\sqrt{(2j+1)(2j+2)}}, \\\label{defcalD}
& {\cal D}_j(|\z_\sR|^2):=\f{(1+2j-|\z_{\sR}|^2)(1+ |\z_\sR|^2)^{d_j} +|\z_\sR|^{4(j+1)}  }{(1+|\z_{\scr R}|^2)^{2j+1} }.
\end{align}
Therefore,
\begin{align}
\lim_{A\rightarrow\infty}\bra{H,t} \wh{e^{i\xi}} \ket{H,t}& = 
e^{i\xi} \sum_j  (2j+3)^\f12(2j+2)^{-\f12}(2j+1)^{-\f12} {e^{-t(j+1)^2} e^{d_j A} } {\cal D}_j(\z_\sL) {\cal D}_j(\z_\sR),
\end{align}
with the desired factor $e^{i\xi}$.
As a consistency check, taking $\z_\sL=0=\z_\sR$ in the above expression we recover the $A\rightarrow\infty$ limit of \Ref{EVxiT1}.

\bigskip
Summarizing, we have seen how the detailed peakedness properties of the heat-kernel coherent states can be exposed computing expectation values.
We have done so for the spin, the fluxes, and the spin-shift operator. The latter captures the component of the holonomy which cannot be reconstructed from knowledge of both left and right-invariant fluxes.  Simple expectation values given directly by the classical labels occur only in the  $A\to\infty$ limit. Also in this limit, the relative uncertainties the norm and directions of the fluxes vanish. The relative uncertainty of the spin-shift operator does not stricly vanish for any finite value of $t$, but this is a feature of compact variables, and the uncertainty is correctly minimized. 
On the other hand, the direction of the fluxes is fuzzy at finite $A$, and cannot be identified with the classical label, and same situation for the spin-shift. The heat-kernel states are chosen to minimize the uncertainty between the creation and annihilation operators of the $\SL(2,\C)$ complex structure, and cannot thus simultaneously minimize the directions. Experience with spin foams and quantum reduced LQG suggests on the other hand the usefulness of working with a family of states all having a clear direction. This can be achieved working with SU(2) coherent states, and at the gauge-invariant level, with the coherent intertwiners of \cite{LS}. In the next Section, we propose a family of coherent states with these properties.

\section{The new family of coherent states}

The new family of coherent states we propose is  
\begin{align}\label{defTGCS}
\Psi^t_{\Om}(g) &= \la \bar g\ket{\Om,t}:= \sum_j d_j^{\f 32}e^{-\f t2j(j+1)} 
\Big(e^{-\f i2 d_j \om}D_{jj}^{(j)}(n^{-1}_\sR g n_\sL) + e^{\f i2 d_j \om}D_{-j,-j}^{(j)}(n^{-1}_\sR g n_\sL)\Big). 
\end{align}
Or in the momentum basis,
\be\label{Omjmn}
\ket{\Om,t} = \sum_{jmn} d_j e^{-\f t2j(j+1)} 
\Big(e^{-\f i2 d_j \om} D^{(j)}_{jm}(n^{-1}_\sR)D_{nj}^{(j)}(n_\sL)  + e^{\f i2 d_j \om}D^{(j)}_{-jm}(n^{-1}_\sR)D_{n,-j}^{(j)}(n_\sL)\Big)
\ket{j,m,n}.
\ee
We refer to \Ref{defTGCS} as twisted geometry coherent states. 
We label them by
$\Om:=(\om,\z_\sL,\z_\sR)$, to distinguish them from the heat-kernel states labelled by $H$, but the two sets of variables parametrize the same phase space, and are thus equivalent. 
The main difference between the heat-kernel (HKCS) and the twisted geometries (TGCS) coherent states \Ref{defTGCS} is that the trace in \Ref{HKCS} has been replaced by the highest and lowest weights alone. That this is possible while still providing an over-complete basis in $L_2[\SU(2)]$ is the main new message of this paper, and will be proved below.

To compare these states
with the large area limit \Ref{HKlimit} of the heat-kernel states,
we rewrite them as
\begin{align} 
\Psi^t_{\Om}(g) & = e^{\f 1{2t}( A -\f t2)^2} e^{-\f i2\om} \sum_j d_j^{\f 32} e^{-\f t2(\f{d_j}2-\f At)^2-i\xi j} D^{(j)}_{jj}(n_\sR^{-1}g  n_\sL)
\\\nn &\quad +e^{\f 1{2t}( A +\f t2)^2} e^{-\f t8+\f i2\om} \sum_j d_j^{\f 32} e^{-\f t2(\f{d_j}2+\f At)^2+i\xi j} D^{(j)}_{-j,-j}(n_\sR^{-1}g  n_\sL).
\end{align}
The second line above is exponentially suppressed for $A\rightarrow\infty$, and we recover  \Ref{HKlimit} up to the prefactor $e^{-\f i2\om}$, and an extra half power of $d_j$.

The norm of the states can be computed to be
\begin{align}\label{norm}
\norm{\Psi_{\Om}^t}^2 &= 2 \sum_j d_j^2 e^{-tj(j+1)}\cosh(d_j A)
=e^{\f t4}(1-4\p_t)\Big(\vartheta_2(iA,e^{-t})+\vartheta_3(iA,e^{-t})-1\Big),
\end{align}
where $\vartheta_i(u,q)$ are Jacobi's theta functions (we use the conventions of Wolfram's Mathematica).
The differential operator acting on the theta functions is a consequence of the $d_j^2$ factor in the sum. 
This factor cannot be removed from the definition of the coherent states, because our resolution of the identity 
would fail otherwise.
The second theta function is needed to take into account the half-integer representations. If one were interested in $L_2[\SO(3)]$ instead of $L_2[\SU(2)]$, the sum in \Ref{defTGCS} would be over the natural numbers alone, and only  $\vartheta_2$ would appear in the norm.

The reason for the presence of both highest and lowest weights  in \Ref{defTGCS} is the $\Z_2$ symmetry of the twisted geometry parametrization of $T^*\SU(2)$ \cite{twigeo}, and the existence of two families of SU(2) coherent states. 
The reader may wonder whether it is possible to take only one term with highest or lowest weights. The answer is yes,
but the resulting states have less compelling properties, like a $\xi$-dependent norm. This will be discussed below in  Section~\ref{SecCos}.

\subsection{Resolution of the identity}
The necessary, and weakest condition that one can require of coherent states, is that they provide a resolution of the identity as an integral over the classical phase space \cite{GazeauBook}.
With our new family of coherent states, the resolution of the identity is given by the following measure,
\begin{equation}\label{mes}
\mathds{1} = e^{-\f t4}\int_{\C P^1}d\m(\z_\sL)\int_{\C P^1}d\m(\z_\sR) \int_{-2\pi}^{2\pi} \frac{d\xi}{4\pi}  \int_{0}^{\infty} \frac{dA}{\sqrt{\pi t}} e^{-\frac{A^2}{t}} \ket{\Om,t}\bra{\Om,t},
\end{equation}
where $d\m(\z)$ is the normalized measure on the sphere.
To prove it, it is sufficient to evaluate
\begin{align}\label{ResId1}
\bra{k,p,q}j,m,n\ra &= e^{-\f t4}d_{j}d_{k} e^{-\f t2C_j-\f t2C_{k}}\int_{\C P^1}d\m(\z_\sL) \int_{\C P^1}d\m(\z_\sR)
\int_{-2\pi}^{2\pi} \frac{d\xi}{4\pi}  \int_{0}^{\infty} \frac{dA}{\sqrt{\pi t}} e^{-\frac{A^2}{t}}
\\\nn &\qquad \times\Big(e^{\f 12 (i\xi+A) d_k } \overline{D^{(k)}_{qk}(n_\sL)} \overline{D_{kp}^{(k)}(n^{-1}_\sR)} 
+ e^{-\f 12 (i \xi+A) d_k}\overline{D^{(k)}_{q,-k}(n_\sL)} \overline{D_{-kp}^{(k)}(n^{-1}_\sR)}  \Big) 
\\\nn &\qquad \times\Big(e^{-\f 12 (i\xi-A) d_{j} } {D^{(j)}_{nj}( n_\sL)} {D_{jm}^{(j)}(n^{-1}_\sR)} +
e^{\f 12 (i\xi-A) d_{j}} {D^{(j)}_{n,-j}( n_\sL)} {D_{-jm}^{(j)}(n^{-1}_\sR)}  \Big), 
\end{align}
and verify that it results in a product of Kronecker deltas.
To proceed, it is simpler to first integrate over $\xi$. There are four possible terms, and the relevant integrals give
\be
 \int_{-2\pi}^{2\pi} \frac{d\xi}{4\pi} e^{-i\xi(j+\f12)} e^{\pm i \xi(k+\f12)} = \left\{\begin{array}{l} \d_{2j,2k} \\ \d_{2j+2,-2k}
 \end{array}\right.
\ee
Notice that when SU(2) is parametrized like in \Ref{twigeo}, the period of $\xi$ is $4\pi$, and this allows one to include the half-integer representations in the delta above. 
We will from now on use the simpler notation $\d_{jk}$ to include the half-integers as well, as customary in $\SU(2)$ theory.
Since the spins are only positive, the $\xi$ integral eliminates the mixed terms between the second and third line of \Ref{ResId1}.
The sphere integrals then result in Kronecker deltas on the magnetic indices, thanks to the resolution of the identity satisfied by the SU(2) coherent states:
\be
 \int \overline{D^{(j)}_{m,-j}(n)} D^{(k)}_{p,-k}(n)= \f 1{d_j}\d_{jk}\d_{mp},
\ee
and similarly the second sphere integral, producing  $\d_{nq}$.
The final integration gives
\begin{align}\label{ResId2}
\bra{j,m,n}k,p,q\ra &= \d_{jk}\d_{mp}\d_{nq}
e^{-\f t4} e^{-tC_j}  \int_{0}^{\infty} \frac{dA}{\sqrt{\pi t}} e^{-\frac{A^2}{t}} \Big(e^{\f 12 A d_j} + e^{-\f 12 A d_j} \Big)
= \d_{jk}\d_{mp}\d_{nq}.
\end{align}

This proves that \Ref{defTGCS} is an overcomplete basis in $L_2[\SU(2)]$.
It is not orthogonal, and the overlap between two states can be 
given a compact explicit form thanks to the properties of the Wigner matrices,
\begin{align}
\langle \Om',t|\Om,t \rangle 
&= e^{-\f t4}\sum_j d_j^2 e^{-tC_j} \f{1}{(1+|\z_\sL'|^2)^j (1+|\z_\sR'|^2)^j  (1+|\z_\sL|^2)^j (1+|\z_\sR|^2)^j }\\\nn
&\quad \times \Big[e^{\f i2 d_j(\bar\om'-\om)}(1+\z_\sL'\bar{\z_\sL})^{2j} (1+\bar \z_\sR' {\z_\sR})^{2j}
+ e^{-\f i2 d_j(\bar\om'-\om)}(1+\bar \z_\sL' {\z_\sL})^{2j} (1+\z_\sR' {\bar\z_\sR})^{2j} \\\nn
&\qquad + e^{\f i2 d_j(\bar\om' +\om)}(-\z_\sL'+ {\z_\sL})^{2j} (-\bar \z_\sR' +{\bar \z_\sR})^{2j}
+ e^{-\f i2 d_j(\bar\om' +\om)}(-\bar\z_\sL'+ {\bar\z_\sL})^{2j} (-\z_\sR' +{ \z_\sR})^{2j}\Big].
\end{align}

\subsection{Expectation values and peakedness properties}

\subsubsection*{Spin }
\begin{figure}[H]
\centering
\includegraphics[width=8cm]{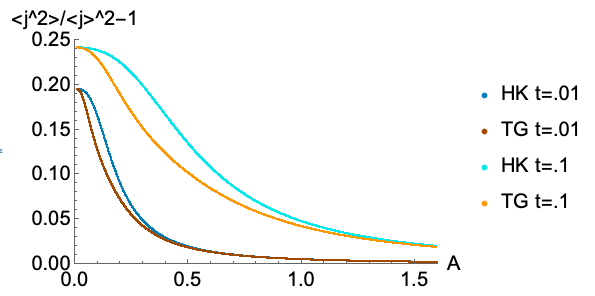}\hspace{.5cm}\includegraphics[width=8cm]{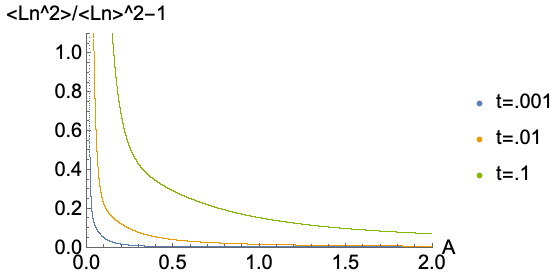}
\caption{\small{Left panel: \emph{Comparison of the spin peakedness for the HK and TG coherent states. The large area behaviour is the same, but the TG coherent states have slightly better spread at finite $A$.} Right panel: \emph{Relative uncertainty of a flux component $i$, with $n_{\sL i}=.3$.
The divergence in zero is due to the vanishing of the expectation value.}}
\label{FigjComp}
}
\end{figure}

The spin expectation value is
\be\label{EVjTG}
\mean{\hat\jmath}:=\f{ \bra{\Om,t}\hat\jmath\ket{\Om,t} }{\norm{\Psi_{\Om}^t}^2} = 
\f{\sum_j j d_j^2 e^{-tj(j+1)}\cosh(d_j A)}{\sum_j d_j^2 e^{-tj(j+1)}\cosh(d_j A)} 
\simeq \left(\f At-\f12+\f1{A}\right), 
\ee
with the NLO approximation valid as before for small $t$ and large $A$. Comparing with \Ref{EVjTLO}, we see that the large $A$ behaviour is the same at leading order, with a different NLO correction. 
For the relative uncertainty we find the same leading order \Ref{RUj} of the heat-kernel states, vanishing in the large area limit. 
The same qualitative picture emerges: the states with the best semiclassical behaviour in the spin are those with large $A$. 
Numerical investigations show that  the spread of the TGCS is slightly better than that of the HKCS at finite $A$, see Fig.~\ref{FigjComp}.

\subsubsection*{Fluxes}

To compute the expectation value of the fluxes, we take advantage of the property \Ref{nLn} of SU(2) coherent states.
A simple calculation shows that
\begin{align}\label{meanL}
&\bra{\Om,t}\hat {\vec L}\ket{\Om,t} = f(A,t)\,\vec n_{\sL}, \qquad f(A,t):= 2 \sum_j j d_j^2 e^{-tj(j+1)}\sinh(d_j A),\\
&\bra{\Om,t} \widehat{L_i^2} \ket{\Om,t} = 2 \sum_j \Big(j^2n_{\sL i}^2+\f j2(1-n_{\sL i}^2)\Big)d_j^2 e^{-tj(j+1)}\cosh(d_j A).
\end{align}
The first expression shows that the TGCS have average direction given exactly by the classical label for any value of $A$. 
This is the main difference with the HKCS. Compare also the simplicity of the above expressions with respect to \Ref{HKLev} and and \Ref{HKLqev}.
The norm of the flux expectation value is however not simply $A$, but reproduces it only for large areas:
\be\label{fluxTGCS}
\mean{\hat {\vec L}}:=\f{ \bra{\Om,t}\hat {\vec L}\ket{\Om,t} }{\norm{\Psi_{\Om}^t}^2} = 
\f{\sum_j j d_j^2 e^{-tj(j+1)}\sinh(d_j A)}{\sum_j d_j^2 e^{-tj(j+1)}\cosh(d_j A)}  \vec n_{\sL}
\simeq \left(\f At-\f12+\f1{A}\right) \vec n_{\scr L}.
\ee
The non-linear dependence on $A$ is the same situation that occurs for the HKCS expectation values, and also for  spin  expectation value \Ref{EVjTG}. As a significative difference, we notice that the spin expectation value is even in $A$ and non-zero also for $A=0$, whereas the flux \Ref{meanL} is odd in $A$ and vanishing for $A=0$.

The leading order in the large area limit of \Ref{fluxTGCS} coincides with \Ref{EVjTLO} obtained with the HKCS, as expected. 
And indeed, also the leading order of the relative uncertainty gives the same value \Ref{RUj}. 
The difference between the two families is to be found at finite $A$, where the TGCS  still identify a clear direction of the fluxes thanks to \Ref{meanL}, unlike the HKCS. The spread around the flux direction direction depends on $A$, and the relative uncertainty only vanishes in the large $A$ limit, see Fig.~\ref{FigjComp}. This is inevitable, after all also the SU(2) coherent states have vanishing relative uncertainty only in the large spin limit. The SU(2) coherent states further minimize the flux uncertainty $\D^2 L_x\D^2L_y\geq\mean{L_z}$. This nice property cannot unfortunately be expected for the full TGCS, because it is spoiled by the spin sums. 

We focused here on the left-invariant fluxes, but the story is the same for the right-invariant fluxes, 
\be
\bra{\Om,t}\hat {\vec R}\ket{\Om,t} \simeq \left(\f At-\f12+\f1{A}\right) \vec n_{\sR},
\ee
and so on.

\subsubsection*{Holonomy peakedness}

As explained earlier, the twisted geometry picture suggests to shift attention from the holonomy-flux operators to the right and left-invariant fluxes plus the twist angle operator, and this is what we consider to study the holonomy peakedness. 
Using \Ref{exiop}, we find
\begin{align}
\label{eqn:horriblexi}
\bra{\Om,t} \wh{e^{i\xi}} \ket{\Om,t}&=e^{-\f t4}\sum_j d_j (2j+3) e^{-t(j+1)^2}
\sum_{m,n} \left( \f{j+m+1}{j+m+2}\f{j+n+1}{j+n+2}\right)^{\tfrac12}
\\\nn &\qquad \times\Big(e^{\f 12 (i\xi+A) d_{j+1} } {D^{(j+1)}_{n+1,j+1}(\bar n_\sL)} {D_{m+1,j+1}^{(j+1)}(n_\sR)} 
+ e^{-\f 12 (i \xi+A) d_{j+1}} {D^{(j+1)}_{n+1,-j-1}(\bar n_\sL)}  {D_{m+1,-j-1}^{(j+1)}(n_\sR)}  \Big) 
\\\nn &\qquad \times\Big(e^{-\f 12 (i\xi-A) d_{j} } {D^{(j)}_{nj}( n_\sL)} {D_{jm}^{(j)}(n^{-1}_\sR)} +
e^{\f 12 (i\xi-A) d_{j}} {D^{(j)}_{n,-j}( n_\sL)} {D_{-jm}^{(j)}(n^{-1}_\sR)}  \Big) \\\nn
&= e^{i\xi} e^{-\f t4}\sum_j \f{(2j+3)}{(2j+2)}  e^{-t(j+1)^2}
\Big[ e^{2(j+1)A}  {\cal D}_j(|\z_\sL|^2) {\cal D}_j(|\z_\sR|^2) + e^{2i\xi+A} \f{(\z_\sL)^{2j}}{(1+|\z_\sL|^2)^{2j+1}} \f{(\z_\sR)^{2j}}{(1+|\z_\sR|^2)^{2j+1}} \Big]
\\\nn &\quad +
e^{-i\xi} e^{-\f t4}\sum_j \f{(2j+3)}{(2j+2)}  e^{-t(j+1)^2}
\Big[ e^{-2(j+1)A}  {\cal D}'_j(|\z_\sL|^2) {\cal D}'_j(|\z_\sR|^2) +  \f{e^{-2i\xi-A} (\bar\z_\sL)^{2j} (\bar\z_\sR)^{2j} }{(1+|\z_\sL|^2)^{2j+1} (1+|\z_\sR|^2)^{2j+1}} \Big],
\end{align}
where we used the definition \Ref{defcalD} and 
\be
{\cal D}'_j(|\z_\sR|^2):=\f{ (-1+d_j|\z_{\sL}|^2)(1+ |\z_\sL|^2)^{d_j} +1  }{(1+|\z_{\sL}|^2)^{2j+1}\, |\z_\sL|^{4}}.
\ee
See Appendix for details.
While we did not achieve a full factorization of the classical value, it is easy to see from the above expression that the second line is 
 exponentially suppressed for large $A$. The normalized expectation value then becomes quickly $e^{i\xi} e^{-\f t4}$.
We can obtain more explicit formulas for the simple case $\z_\sL=0=\z_\sR$, 
\begin{align}
\bra{e^{\om\t_3},t} \wh{e^{i\xi}} \ket{e^{\om\t_3},t}&= e^{i\xi} e^{-\f t4}\sum_j \f{(2j+1)^2(2j+3)}{(2j+2)}  e^{-t(j+1)^2} e^{(2j+2)A},
\end{align}
\be
\mean{\wh{e^{i\xi}} }
=e^{i\xi} \f{\sum_j \f{(2j+1)^2(2j+3)}{(2j+2)}  e^{-t(j+1)^2} e^{(2j+2)A}}{2\sum_j d_j^2 e^{-tj(j+1)}\cosh(d_j A)} 
\stackrel{A\mapsto\infty}{\longrightarrow} e^{i\xi} e^{-\f t4}.
\ee
In this simpler case, the variance is
\begin{align}
\bra{\Om,t} \wh{e^{2i\xi}} \ket{\Om,t}&= e^{2i\xi} e^{-\f t4}\sum_j \f{(2j+1)^2(2j+3)(2j+5)}{(2j+2)(2j+4)}  e^{-t(j^2+3j+3)} e^{(2j+3)A}
\stackrel{A\mapsto\infty}{\longrightarrow} e^{2i\xi} e^{-t}.
\end{align}
We recover the same peakedness properties of the heat-kernel coherent states, included the typical shift in the expectation values of the angle operators.

\begin{figure}[H]
\centering
\includegraphics[width=8cm]{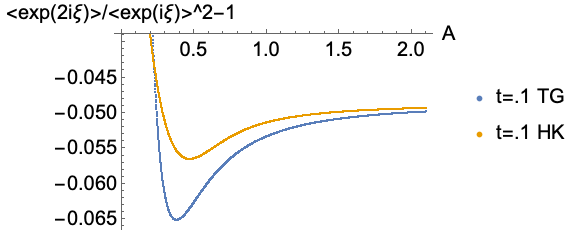}
\caption{\small{\emph{Comparison of the spread of $\wh{e^{i\xi}}$ for the HK and TG coherent states. The large area behaviour is the same, but the HK coherent states have slightly better spread at finite $A$.}}
\label{FigComparison}
}
\end{figure}

These expressions are simpler than the corresponding ones for the HKCS, see e.g. \Ref{EVxiT1}, and with the same large $A$ limit.
On the other hand, the states are slightly more spread than the HKCS for finite $A$, as shown in Fig.~\ref{FigComparison}. This is a price to pay for having better peakedness properties in the directions. Together of course with the loss of minimal uncertainty between the annihilation operator $\hat H$ and its adjoint.

\subsection{Simpler cosine states}\label{SecCos}

It is possible to consider a simplified family given by only one set of extremal weights, e.g. the lowest,
\be\label{defCSs}
\Psi{}^t_{\Om\rm c}(g) :=2 \sum_j d_j^{\f 32}e^{-\f t2j(j+1)} \cos\left(\tfrac 12 d_j \om\right) D_{-j-j}^{(j)}(n^{-1}_\sR g n_{\sL}).
\ee
The necessary modification is to use a cosine of $d_j\om$ instead of a single exponential. We label this choice with $\Om$c to distinguish it from the symmetric choice \Ref{defTGCS}, with the c in reference to the cosine. 
This is also a good family of coherent states, and provides a resolution of the identity with exactly the same measure \Ref{mes}.
It has the nice property of being a holomorphic function in $(\om,\z_{\sL},\bar\z_{\sR})$,\footnote{A version holomorphic in $(\om,\z_{\sL},\z_{\sR})$ is obtained taking mixed highest/lowest weights $D_{j-j}$.}, even though unfortunately this does not lead to the possibility of constructing a new holomorphic representation for the holonomy-flux algebra (more on this below).
It has similar peakedness properties to the symmetric family \Ref{defTGCS}, 
but a somewhat unpalatable $\xi$-dependence of norm and flux expectation values. For the norm of the states, we have
\be
\norm{ \Psi{}^{t}_{\Om\rm c}}^2 = 2\sum_j d_j^2 e^{-tj(j+1)}(\cosh(d_j A)+\cos(d_j\xi)).
\ee
The dependence on $\xi$ fades in the large $A$ limit only, and the elegant connection to the theta functions is lost.
For the flux expectation values, we find 
\be
\mean{\hat{\vec L}} = \f{\sum_j j d_j^2 e^{-tj(j+1)}\Big(\cosh(d_j A)+\cos(d_j\xi)\Big)}{\sum_j d_j^2 e^{-tj(j+1)}\Big(\cosh(d_j A)+\cos(d_j\xi)\Big)} \, \vec n_{\sL}.
\ee
We recover the nice property of a simple direction given by the classical label, but the proportionality factor depends again on $\xi$, and not only on $A$.
For these reasons, they seem to us a less elegant option. 
On the other hand, certain mathematical manipulations can be simpler with these states thanks to the presence of a single extremal weight.
With these considerations upfront, both choices \Ref{defTGCS} and \Ref{defCSs} provide families of coherent states.

\section{Further properties of the new coherent states}

\subsection{$\Z_2$ symmetry of twisted geometries}

The rationale for the new coherent states is to take inspiration from the twisted geometry parametrization \Ref{twigeo}.
To begin with, we can peak the fluxes on the classical labels $n_{\sL,\sR}$ using linear superpositions in the magnetic numbers with the coefficients of the SU(2) coherent states, 
\be
\sum_{jmn} w_j D^{(j)}_{\pm j,m}(n^{-1}_\sR)D_{n,\pm j}^{(j)}(n_\sL)\,\ket{j,m,n}.
\ee
The spin weights $w_j$ can then be chosen to introduce a peakedness in the area-angle pair $(A,\xi)$. 
To find the right weights, we look at the area-angle part of the algebra. Taken by itself, this has the same Poisson structure of the pair of action-angle  variables for the harmonic oscillator. One way to construct coherent states for the action-angle variables is to start from the coherent states for a particle on the circle, 
\be
\ket{\om,t} = \sum_{n\in\Z} e^{-\f t2 n^2-in\om }\ket{n} \in L_2[S^1],
\ee
and reduce them by a $\Z_2$ symmetry mapping the negative modes of the particle into positive values of the harmonic oscillator action.
This reduction results in trading $\exp(-in\om)$ with $\cos(n\om)$.
To apply this procedure in our case, two  adaptations are needed. First, we start with a particle on the double cover of the circle, to allow for half-integer representations. Second, we shift the particle's momentum to $j+1/2$ so to have the Casimir as Gaussian factor. 
Accordingly, we consider as starting ansatz the states
\be\label{halfstates}
\ket{\Om,t}_{\pm}:=\sum_{jmn} d_j e^{-\f t2 j(j+1)}
e^{\mp\f i2 d_j \om} D^{(j)}_{\pm j,m}(n^{-1}_\sR)D_{n,\pm j}^{(j)}(n_\sL)\,\ket{j,m,n},
\ee
with $j\in\Z/2$. Both choices of sign are equally good, and we have a single exponential and extremal weight.
The $\Z_2$ symmetry is not implemented yet:
As such, these states are the analogue of the particle on the circle, and would provide a resolution of the identity on an auxiliary Hilbert space with $j\in\Z/2$, and double cover phase space
$A\in\R,\ \xi\in[-4\pi,4\pi), \ \z_\sL\in\C P^1,  \ \z_\sR\in\C P^1$, 
given by
\begin{equation}\label{mesdouble}
\mathds{1}_{\rm aux} = e^{-\f t4}\int_{\C P^1}d\m(\z_\sL)\int_{\C P^1}d\m(\z_\sR) \int_{-4\pi}^{4\pi} \frac{d\xi}{8\pi}  \int_{-\infty}^{\infty} \frac{dA}{\sqrt{\pi t}} e^{-\frac{A^2}{t}} \ket{\Om,t}_\pm\bra{\Om,t}_\pm.
\end{equation}
To restrict the integration domain to the proper phase space, there are two simple options that manifest themselves. 
First, we could modify the Gaussian factor, introducing an Erf function as discussed earlier for the Gaussian states \Ref{PsiG}. 
Observing that
\begin{align}\label{ResId2}
e^{-\f t4}e^{-t C_j}\int_{0}^{\infty} \frac{dA}{\sqrt{\pi t}} e^{-\frac{A^2}{t}} e^{-A d_j} &= \f12 \bigg(1+\textrm{Erf}\Big(\f{d_j}2\sqrt{t}\Big)\bigg),
\end{align}
The correct resolution of the identity is  satisfied if we modify \Ref{halfstates} with an additional Erf weight,
\begin{align}\label{Aintpos}
& \ket{\Om,t}^{\scr Erf}_{\pm}:=\sum_{jmn} d_j \sqrt 2 \bigg(1+\textrm{Erf}\Big(\f{d_j}2\sqrt{t}\Big)\bigg)^{-\f12} e^{-\f t2 j(j+1)}
e^{\mp\f i2 d_j \om} D^{(j)}_{\pm j,m}(n^{-1}_\sR)D_{n,\pm j}^{(j)}(n_\sL)\,\ket{j,m,n},\\
& \mathds{1} = e^{-\f t4}\int_{\C P^1}d\m(\z_\sL)\int_{\C P^1}d\m(\z_\sR) \int_{-2\pi}^{2\pi} \frac{d\xi}{4\pi}  \int_{0}^{\infty} \frac{dA}{\sqrt{\pi t}} e^{-\frac{A^2}{t}} \ket{\Om,t}^{\scr Erf}_\pm\bra{\Om,t}^{\scr Erf}_\pm.
\end{align}
This is the same type of approach that one could use for the Gaussian states \Ref{PsiG}. The result is a family of coherent states with rather complicated expressions for the expectation values. 

The second option is to select even functions of the phase space. The map between the double cover phase space and $T^*\SU(2)$ has the following $\Z_2$ symmetry,
\be\label{Z2map}
(\om,n_{\sL},n_{\sR}) \mapsto (-\om,n_{\sL}\eps,n_{\sR}\eps).
\ee 
The action on the Hopf section has the effect of switching highest and lowest weights.
If we restrict the ansatz \Ref{halfstates} requiring even functions under \Ref{Z2map} we obtain the twisted geometries coherent states \Ref{defTGCS},
\be
\ket{\Om,t}=\ket{\Om,t}_{-}+\ket{\Om,t}_{+}.
\ee
For the simpler cosine states \Ref{defCSs}, we require even functions under the map $\om\mapsto -\om$. 
The full map \Ref{Z2map} does not leave the cosine states invariant, but trades them for an equivalent set. 
Therefore both families implement the $\Z_2$ symmetry.

\subsection{On holomorphic representations}

A beautiful property of the HKCS is to define a holomorphic representation for the holonomy-flux operators, based on the $\SL(2,\C)$ complex structure.
In this representation, the annihilation operator $a:=\hat H$ and its adjoint act as differentiation and multiplication by a triple of complex variables.
These are in turns functions of the original holonomy-flux operators.
A natural question for us is whether the new family of TGCS offers a new holomorphic representation, possibly related to the holonomy-flux operators in a simpler way than the creation and annihilation operators. After all, the parametrization \Ref{twigeo} is suggestive that there could be a holomorphic representation based on the complex structures of $S^2$ and $T^*S^1$, as opposed to that of $\SL(2,\C)$. 
The answer appears however to be negative.

The simplest attempt to construct such a new holomorphic representation would be to start from the known holomorphic representations of SU(2) and the particle on the circle. Namely, to consider first the auxiliary Hilbert space discussed above, the one with $j\in\Z/2$.
In this auxiliary space, the states \Ref{halfstates} provide indeed a holomorphic representation, with
\be\label{jholo}
\hat\jmath = \mp i\p_\om-\f12,
\ee
depending on the $\pm1$ choice in \Ref{halfstates}, and the fluxes acting as the known holomorphic representation given by the $\SU(2)$ coherent states \cite{Perelomov}.
This representation however fails for the states \Ref{defTGCS} or \Ref{defCSs} because it is incompatible with the $\Z_2$ symmetry they implement:
In a very obvious way, the derivative does not preserve the cosine choice of the simpler states, nor the structure of the full states.

One may try more complicated ans\"atze than \Ref{jholo}, but we believe there is a deeper stumbling block. The holomorphic representation provided by the HKCS is guaranteed by the complexifier construction. As far as we understand, this requires global complex Darboux coordinates, which are (non-trivially) provided by the $\SL(2,\C)$ complex structure. On the other hand,
even if the parametrization \Ref{twigeo} splits the $T^*\SU(2)$ symplectic potential into three terms corresponding to complex manifolds, the three complex variables $(\om,\z_{\sL},\z_{\sR})$ do not provide global Darboux coordinates. Therefore the states obtained are not of the complexifier type, and one cannot obtain a holomorphic representation that way. 

An alternative and elegant mathematical way to construct holomorphic representations is to use geometric quantization. In this approach, one uses builds a representation associated to a classical complex polarization taking the symplectic potential as connection, and requiring covariant constancy of the wave functions. For a simple Darboux symplectic potential, this translates into holomorphicity of the wave-functions.
But the symplectic potential in the twisted geometry variables is not diagonal \cite{twigeo}. Therefore imposing this strategy will indeed not lead to simple holomorphic functions. We leave the investigation of this approach to future work.

\subsection{Annihilation operator eigenvalue equations}

By the way we constructed the new coherent states, there is no guarantee that they are eigenvectors of an annihilation operator. And indeed this appears not to be the case. We consider the following shift operators,
\begin{align}
& \hS\,|j,m,n\ra=\f{1}{2j(2j+1)}\, \sqrt{(j+m)(j+m-1)(j+n)(j+n-1)} \,|j-1,m-1,n-1\ra, \\
& \hT\,|j,m,n\ra=\f{1}{2j(2j+1)}\,\sqrt{(j-m)(j-m-1)(j-n)(j-n-1)}\,|j-1,m+1,n+1\ra.
\end{align}
Since they decrease the spin $j$, they can be interpreted as annihilation operators while their adjoint operators are interpreted as creation operators.
The auxiliary states \Ref{halfstates} satisfy the following finite difference eigenvalue equations:
\begin{align}
& \hS\,e^{t(\hat{\jmath}+1)}\,\ket{\Om,t}_{-} = A\,\ket{\Om,t}_{-}, &&
\hT\,e^{t(\hat{\jmath}+1)}\,\ket{\Om,t}_{-}= \bar{B}\,\ket{\Om,t}_{-}, \\
& \hS\,e^{t(\hat{\jmath}+1)}\,\ket{\Om,t}_{+} = B\,\ket{\Om,t}_{+}, &&
\hT\,e^{t(\hat{\jmath}+1)}\,\ket{\Om,t}_{+} = \bar{A}\,\ket{\Om,t}_{+},
\end{align}
where
\be 
A:=e^{i\om}\f{\z_\sL^{2}\bar\z_\sR^{2}}{\sqrt{(1+|\z_\sL|^2)(1+|\z_\sR|^2)}},\qquad
B:=e^{-i\om}\f{1}{\sqrt{(1+|\z_\sL|^2)(1+|\z_\sR|^2)}}.
\ee
Hence these states are eigenvectors of $\hat S$ and $\hat T$.
From these formulas and a little algebra, we find that the $\Z_{2}$-invariant states satisfy
\be
\left[\hS+e^{-2i\om}\f{e^{+2i\om}\z_\sL^{2}\bar\z_\sR^{2}-1}{e^{-2i\om}\bar\z_\sL^{2}\z_\sR^{2}-1}
\hT
\right]\,e^{t\hat{\jmath}}\,\ket{\Om,t}=\f{e^{-t}e^{-i\om}}{e^{-2i\om}\bar\z_\sL^{2}\z_\sR^{2}-1}
\f{|\z_\sL^{2}\bar\z_\sR^{2}|^2-1}{\sqrt{(1+|\z_\sL|^2)(1+|\z_\sR|^2)}}\,\ket{\Om,t}.
\ee
Notice the phase-space dependence on the left-hand side: even if this equation may look at first sight like an eigenvalue equation for an annihilation operator, each state in the family requires a different operator, and therefore the members of the family are not eigenvectors of the same operators. 
It is possible to consider different annihilation operators acting differently on the magnetic labels, but we were not able to find any whose eigenvectors span the new family of coherent states.
Just as for the holomorphic representation, being a family of eigenvectors of annihilation operator is a special property guaranteed by the complexifier construction, but not by our generalized construction of coherent states.

\section{Comparison with the spinorial coherent states}
\label{app:spinorialCS}

We have focused our discussion on the comparison between our new twisted geometry coherent states with the original heat kernel states. 
But  another family of coherent states has also  appeared in the literature on spin networks and spin foam amplitudes for loop quantum gravity. For completeness, we report the comparison with those spinorial coherent states as well. They turn out to be defined by a Poisson weight for the spins instead of the Gaussian probability distribution used in both the HK and TG states.

In the papers on the spinorial representation of loop quantum gravity \cite{Borja:2010rc,Dupuis:2010iq,Dupuis:2011fz,Livine:2011gp,Livine:2013zha}, the following spinorial coherent states were introduced, as natural  functions over the $\SU(2)$ group:
\be
\Psi_{(z_\sL,z_\sR)}(g)=\bra{\bar g }z_\sL,z_\sR\ra:=e^{\la z_{\sR}|g|z_{\sL}\ra}
=\sum_{j} \f{\la z_{\sR}|g|z_{\sL}\ra^{2j}}{(2j)!}
=\sum_{j} \f{\la j,z_{\sR}|g|j,z_{\sL}\ra}{(2j)!},
\ee
with the $\SU(2)$ spinorial coherent states defined as:
\be
\ket{z}=\left(\begin{array}{c}z^{0}\\z^{1}\end{array}\right),\qquad
|j,z\ra={\norm z^{2j}}\,g(z)\,|j,j\ra,\qquad
g(z)=\f1{\norm z}\,\mat{z^{0}}{-\bar{z}^{1}}{z^{1}}{\bar{z}^{0}}\in\SU(2).
\ee
These plane waves $e^{\la z_{\sR}|g|z_{\sL}\ra}$ provide a  resolution of the identity with the Gaussian measure on the spinors \cite{Dupuis:2010iq,Dupuis:2011fz}. They admit a straightforward extension to intertwiner states \cite{Dupuis:2011fz,Bonzom:2012bn} and lead to  interesting applications to spinfoam amplitudes \cite{Hnybida:2014mwa,Banburski:2014cwa,Hnybida:2015ioa}.

The mapping to the twisted geometry parametrization was given in \cite{twigeo2}, see Appendix~\ref{AppA}, and results from a factorization of the $\SU(2)$  matrix elements:
\be
g(z)=
e^{i\textrm{arg}(z_{0})\sigma_{3}}\,n(\z)
\quad\textrm{with}\quad
\z=-\f{\bar{z}^{1}}{z^{0}}
\,.
\ee
This allows to write the plane-wave as:
\be
\label{eqn:spinorialCS}
e^{\la z_{\sR}|g|z_{\sL}\ra}
=
\sum_{j}
\f{(\norm{z_{\sL}}\norm{z_{\sR}})^{2j}}{(2j)!}
\,e^{2ij(\textrm{arg}\,z^0_{\sL}-\textrm{arg}\,z^0_{\sR})}\,D^{(j)}_{jj}(n_{\sR}^{-1}gn_{\sL}),
\ee
where the twisted geometry parameters are thus given in terms of the spinors by $2A=\norm{z_{\sL}}^2=\norm{ z_{\sR}}^2$ and $\xi=2(\textrm{arg}\,z^0_{\sR}-\textrm{arg}\,z^0{\sL})$.

Comparing to the TGCS ansatz, we see that the Gaussian factor is  replaced here by a simple Poisson distribution. The peakedness properties in the normals and holonomy are similar. The main difference is that the spinorial coherent state ansatz does not include any modifiable squeezing parameter $t$, and the width of the state is determined by the area label alone \cite{Dupuis:2010iq,Dupuis:2011fz}.

Let us look at the properties of those states in more details.
Translating the $\SU(2)$ plane wave ansatz given above into the ket notation of coherent states, the spinorial coherent state is defined as:
\bea
|z_\sL,z_\sR\ra&=&
\sum_{j,m,n}
d_{j}^{-\f12}\f{(2A)^{2j}}{(2j)!}
\,e^{-ij\xi}
\,\overline{D^{(j)}_{m,j}(n_{\sR})}\,D^{(j)}_{n,j}(n_{\sL})
\,|j,m,n\ra
\\
&=&
\sum_{j,m,n}
d_{j}^{-\f12}\,\f{(2A)^{2j}}{{\sqrt{(j+m)!(j-m)!(j+n)!(j-n)!}}}
\,e^{-ij\xi}
\,\f{(-\bar\z_\sL)^{j-n}}{(1+|\z_\sL|^2)^{j}}\f{(-\z_\sR)^{j-m}}{(1+|\z_\sR|^2)^{j}}
\,|j,m,n\ra.
\nn
\eea
The norm of this state is:
\be
\la z_\sL,z_\sR|z_\sL,z_\sR\ra
=
\sum_{j\in\N/2}\f{(2A)^{4j}}{(2j)!(2j+1)!}
=
\f{I_{1}(4A)}{2A}\,,
\ee
in terms of the modified Bessel function $I_{1}$.
The probability distribution for the spin $j$ is Poisson-like:
\be
\la z_\sL,z_\sR|\,{\cal P}(\hat{\jmath})\,|z_\sL,z_\sR\ra
=
\sum_{j\in\N/2}{\cal P}(j)\,\f{(2A)^{4j}}{(2j)!(2j+1)!},
\ee
for an arbitrary polynomial ${\cal P}$ in the spin $j$.
The weight factors $\f{(2A)^{4j}}{(2j)!(2j+1)!}$ are peaked on $j\sim A$, which is easy to check for large $A$ using the Stirling approximation for the factorials.
Moreover, including the norm factor, the expectation value of the spin and its spread are given by Bessel functions:
\begin{align}
&\la \hat\jmath\ra:=\f{\la z_\sL,z_\sR|\,\hat{\jmath}\,|z_\sL,z_\sR\ra}{\la z_\sL,z_\sR|z_\sL,z_\sR\ra}
=A\f{I_{2}(4A)}{I_{1}(4A)}
\underset{A\rightarrow\infty}\sim A-\f38, \\
& \la \hat\jmath^2\ra:=\f{\la z_\sL,z_\sR|\,\hat{\jmath}^2\,|z_\sL,z_\sR\ra}{\la z_\sL,z_\sR|z_\sL,z_\sR\ra}
=\f A 2\,\f{I_{3}(4A)+2AI_{2}(4A)}{I_{1}(4A)}\underset{A\rightarrow\infty}\sim A^2+\f A8 -\f{991}{2048},\\
& \D^2\hat\jmath 
\underset{A\rightarrow\infty}\sim \f {7}{8A}\qquad\underset{A\rightarrow\infty}\rightarrow 0.
\end{align}
Now, for the twist angle, we compute the expectation of the exponentiated operator:
\be
\exiop \ket{j,m,n} = \f{(j+m+1)^{1/2}}{(j+m+2)^{1/2}} \f{(j+n+1)^{1/2}}{(j+n+2)^{1/2}} \ket{j+1,m+1,n+1},
\ee
\be
\la z_\sL,z_\sR|\,\exiop\,|z_\sL,z_\sR\ra
=
e^{i\xi}\,
\sum_{j\ge 0}(d_{j}d_{j+1})^{-\f12}(2A)^{4j+2}
\cD_{j}(|\z_{\sL}|^2)\cD_{j}(|\z_{\sR}|^2)
\ee
with the auxiliary function
\be
\cD_{j}(x)
=\f1{(1+x)^{2j+1}}\,\sum_{m}\f{x^{j-m}}{(j-m)!(j+m)!(j+m+2)}
=\f{x^{2j+2}+(d_{j}-x)(1+x)^{2j+1}}{(2j+2)!(1+x)^{2j+1}}.
\ee
The square-root factor $(d_{j}d_{j+1})^{\f12}$ forbids an exact re-summation of this series. This suggests that a slightly different choice of ordering for the translation generator $\exiop$ might be more suitable. Putting this consideration aside, for large spins $j\gg x$,  approximating the factor $(d_{j}d_{j+1})^{\f12}\sim d_{j+\f12}=(2j+2)$, the series simplifies to:
\be
\la z_\sL,z_\sR|\,\exiop\,|z_\sL,z_\sR\ra
\sim
e^{i\xi}\
\sum_{j}\f{(2A)^{4j+2}}{(2j+1)!(2j+2)!}
=
e^{i\xi}\,\left[
\f{I_{1}(4A)}{2A}-1
\right]
\,,
\nn
\ee
such that one gets the expected result at large area $A$:
\be
\la \exiop\ra=\f{\la z_\sL,z_\sR|\,\exiop\,|z_\sL,z_\sR\ra}{\la z_\sL,z_\sR|z_\sL,z_\sR\ra}
\underset{A\rightarrow\infty}\sim
e^{i\xi}
\,.
\ee

\medskip

As for the resolution of the identity, the calculation is easily adaptable from the computation done for the heat kernel coherent state and the $\Z_{2}$-symmetrized coherent state. It is slightly different from the resolution of the identity for spinorial coherent states proposed in \cite{Dupuis:2010iq,Dupuis:2011fz,Livine:2011gp} in terms of complex integrals over the spinors $z_{\sL}$ and $z_{\sR}$, since here we separate the norm of the spinors, from their phases and their 3d directions and further require that their norm be equal. Nevertheless, we get
\footnote{We use the integral identity $\int dA \,K_{1}(4A)A^{2n}=n!(n-1)!2^{-2n-3}$ for an integer $n$.}
:
\be
\mathds{1} = \int_{\C P^1}d\m(\z_\sL)\int_{\C P^1}d\m(\z_\sR) \int_{-2\pi}^{2\pi} \frac{d\xi}{4\pi} 
\int_{0}^{\infty} {dA}\,2^5A^2K_{1}(4A)
|z_\sL,z_\sR\ra\la z_\sL,z_\sR|\,,
\ee
in terms of the modified Bessel
\footnote{This Bessel measure can be simply recovered from a double exponential measure, writing $A=\sqrt{A_{\sL}A_{\sR}}$ with $A_{L,R}$ respectively the half squared-norm of the spinor $z_{L,R}$, and integration measure $e^{-A_{\sL}}e^{-A_{\sR}}\,dA_{\sL}\,dA_{\sR}$.}
function $K_{1}$. Note that in the asymptotic large $A$ regime, this measure factor decreases exponentially as $K_{1}(4A)\propto e^{{-4A}}/\sqrt{A}$.

\medskip

As for the annihilation operator eigenstate equation, we introduce a slightly different normalization\footnotemark{} for the shift operator:
\be
\widehat{\tilde{S}}\,|j,m,n\ra=
\sqrt{\f{(2j+1)}{(2j-1)}}\,
\sqrt{(j+m)(j+m-1)(j+n)(j+n-1)}
\,|j-1,m-1,n-1\ra
\,,
\ee
which is such that:
\be
\widehat{\tilde{S}}\,\ket{z_\sL,z_\sR}=
\f{(2A)^2e^{-i\xi}}{(1+|\z_{\sL}|^2)(1+|\z_{\sR}|^2)}
\ket{z_\sL,z_\sR}
\,,
\ee
thereby providing the spinorial coherent states with a holomorphic representation as eigenstates of an annihilation operator.
\footnotetext{
The difference of normalization with the TGCS comes from the $d_{j}$ factors entering the defining series over the spin $j$.
}

All in all, the spinorial coherent states have all the good properties of coherent states and are elegantly resummed as the plane-waves $\exp{\la z_{\sR}|g|z_{\sL}\ra}$ on the Lie group $\SU(2)$. They can be very efficient for dealing with spin network evaluations and spin foam amplitudes, but they lack the flexibility of a variable width parameter, such as $t$, which is accommodated by the heat kernel and twisted geometry coherent states. It is likely possible to re-introduce such a variable width parameter through a squeezing operation, which would likely complicate the series in spin in a similar way than the HKCS and TGCS.

\section{Gauge-invariant states and coherent intertwiners}

An advantages of the twisted geometry parametrization is the localization of  variables at the nodes, which allows one to solve explicitly the Gauss constraint, and to find a complete set of fully gauge-invariant observables. This was pointed out in \cite{twigeo}, and the explicit reduction developed in \cite{EteraHoloQT,IoPoly} for the node variables, 
and \cite{FreidelJeff13} for the links (see also \cite{DittrichRyan,IoFabio}). 
The number of reduced variables on a generic closed graph is
\be
6L-6N = 2L + 2\sum_n({\rm val}_n-3).
\ee
The node variables, two for 4-valent nodes, can be taken to be dihedral angles, or Kapovich-Millson variables, or complex cross ratios $Z_n$. 
For the links, the area variables $A_l$ are gauge-invariant, but the twist angles $\xi_l$ are not. In fact, they are pure gauge, since any rotation around the $z$ axis of the source or target node changes $\xi_l$ arbitrarily. A gauge-invariant twist angle $\bar\xi^l_{ij}$, depending on the link and a choice of dual edge, can be defined considering the scalar product between the two vectors associated to the same edge in two adjacent polyhedra, that is \cite{DittrichRyan,FreidelJeff13,IoFabio}
\be\label{defxg}
\cos\bar\xi^l_{ij}:= \frac{\vec{n}_{\sL}\times\vec{n}_{i}\cdot R(g_l) \big(\vec{n}_{j}\times\vec{n}_{\sL}\big)}{\norm{\vec{n}_{\sL}\times\vec{n}_{i}} \, \norm{\vec{n}_{j}\times\vec{n}_{\sL}}}.
\ee
Notice the presence of the link holonomy, via the rotation acting on the second edge vector $\vec{n}_{j}\times\vec{n}_{\sL}$.
It parallel transports the vectors in the same frame and guarantees the gauge-invariance of this expression.
Gauge-invariant quantities including the holonomies are typically built using Wilson loops and spin networks. The advantage of this alternative choice is to be more local: 
The gauge-invariant twist angle only requires variables on one link and two half-links connected to it. This description also makes it trivial to solve the problem of redundancy of the loop variables: it suffices to take one choice of $\bar\xi^l_{jk}$ per link, and with this choice done,
\be\label{redvar}
(A_l,\bar\xi{}^l_{jk}, Z_n)
\ee
is a complete and non-redundant basis of observables.
The subset of twisted geometry corresponding to conformal twisted geometries is identified by the conditions $\bar\xi^l_{jk}=\bar\xi^l_{mn}$,
namely the gauge-invariant twist angle is independent of the choice of dual edge. The gauge-invariant twist angle can then be precisely related to the extrinsic curvature. 
For a triangulation, this subset corresponds to a (polyhedral) Regge geometry, but for a generic cellular decomposition, one needs additional shape matching conditions to reduce a conformal twisted geometry to a Regge geometry. See \cite{IoFabio,IoMiklos,Dona:2020yao} for details.

The localisation at the nodes is a convenient feature also of the new family of coherent states. It is expressed by the factorization in terms of highest and lowest weights, without sums over the magnetic indices as in the case of the heat-kernel states. This structure  allows one to immediately write the projection on the gauge invariant states in terms of coherent intertwiners \cite{LS}, for all values of $A$. Consider first the simpler family of cosine states \Ref{defCSs}. Group averaging gives
\be\label{GIcs1}
\ket{\G,{\Om}_l,t}_{\rm c}= \sum_{j_l,i_n} \prod_l d^{\f12}_{j_l} e^{-\f t2{j_l(j_l+1)}} 2 \cos\left(\tfrac 12 d_{j_l} \om_l\right)
\prod_n c_{i_n}(\{n_l\}) \ket{\G,j_l,i_n},
\ee
a linear combination of spin networks weighted by a Gaussian factor -- which can be made explicit using \Ref{Gfactor} -- and the coherent intertwiners.
The explicit form of the coherent intertwiner coefficients $c_{i_n}(\{n_l\})$ is reported in Appendix~\ref{AppCI}, and depends on the orientation of the graph. With the parametrization \Ref{twigeo}, we always associate the left labels $n_{\sL}$ to the sources of a link, and the right labels $n_{\sR}$ to the targets. The argument of the coefficient is then $n_{\sL}$ for the outgoing links and $n_{\sR}$ for the incoming links. 

With the symmetric family \Ref{defTGCS},  group averaging gives
\be\label{GIcs2}
\ket{\G,{\Om}_l,t}= \sum_{j_l,i_n} \prod_l d^{\f12}_{j_l} e^{-\f t2{j_l(j_l+1)}}
\sum_{\eps_l=\pm 1} \prod_l e^{-\f i2 \eps_l d_{j_l} \om_l} \prod_n c^{\s_n}_{i_n}(\{n_l\}) \ket{\G,j_l,i_n}.
\ee
Here $\s_n=\{\eps_l\}$ is the set of signs for all links at the node $n$. The coefficients are the same as before, but now the explicit form is not determined by the orientation of the graph alone, but also by whether one is looking at the lowest or highest weight term. Notice in fact that 
replacing a minimal weight with a maximal weight has exactly the same effect in the coefficient of the coherent intertwiners as changing the orientation of a link. The argument of the coefficient is on the other hand always $n_{\sL}$ for the outgoing links and $n_{\sR}$ for the incoming links. 
In other words, the relative orientation of the normal determined by using the lowest or highest spin affects the form of the coefficient, whereas the label is always the one determined by the graph orientation. With the simpler cosine states not only the attribution of the label is fixed once and for all by the graph orientation, but also the form of the coefficient.

The additional sums, as well as the disconnect between the form of the coefficient $c_{i_n}$ and the orientation of the graph, may look like unnecessary complications of the symmetric family. However, recall that the norm of the coherent intertwiners is exponentially suppressed in the large spin limit unless the closure condition is satisfied \cite{LS}. For the complete states, a large spin limit is induced by a large $A$ limit.
For a node with all links outgoing, the closure condition applied to the labels of \Ref{GIcs2} gives
\be
\sum_{l\in n} j_l \eps_l \vec n_{l\sL} = 0.
\ee
But for a given state the labels are fixed: Therefore {only two} (opposite) sets of $\eps_l$ are not exponentially suppressed in the large $A$ limit. A similar situation occurs for any type of orientation, so that a single sign freedom for the whole graph is enhanced. 
We conclude that in the large $A$ limit, the symmetric states give a reduced state which has the simple form \Ref{GIcs1}, up to exponentially small correction. 

The states \Ref{GIcs1} and \Ref{GIcs2} provide a family of coherent states for the gauge-invariant Hilbert space of loop quantum gravity. 
It is expressed in terms of coherent intertwiners, proving that the latter are a natural and convenient basis for full loop quantum gravity, and not just for the Hilbert space of a single node. 
It can be compared with the corresponding gauge-invariant expressions for the HKCS studied in \cite{Bahr:2007xn}.

The presence of states with exponentially-suppressed norm in these two families (those with the normals not closing) is a common result of quantum projections, which leave us with an over-redundant family of gauge-invariant coherent states labelled by non-gauge-invariant classical labels. 
 As a consequence, the resolution of the identity over the gauge-invariant Hilbert space using \Ref{GIcs1} or \Ref{GIcs2} counts infinitely many times the same state.
For a single coherent intertwiner, 
a further reduction of the classical labels to gauge-invariant ones was obtained in \cite{EteraHoloQT} for a four-valent node, and in \cite{Bonzom:2012bn} for a general node. The result is a family of reduced coherent intertwiners labelled only by gauge-invariant variables, like the cross ratios for instance, and providing a minimal resolution of the identity without over-redundancies.
The interesting question that we leave open for future work is to do the same at the level of the full graph Hilbert space, and reduce the gauge-invariant states  \Ref{GIcs1} or \Ref{GIcs2} so that they depend only on the gauge-invariant variables \Ref{redvar}.
We expect that doing so will also eliminate the $\eps_l$ sums of the symmetric states, providing simpler expressions.

\section{Conclusions}

We have shown that it is possible to have coherent states for LQG that have always a clear direction of the fluxes, while keeping good peakedness properties in the fluxes and holonomy components. For the holonomy, we stressed the usefulness of defining simpler shifts operators using the harmonic oscillator representation defined by the twisted geometry and spinorial parametrization. A direct numerical study of peakedness shows that the new TGCS are more peaked in the flux than the HKCS, but less peaked in the holonomy component studied. The two key properties of the TGCS are the expectation value of the fluxes, with direction exactly given by the classical label, see \Ref{meanL}, and their relation to coherent intertwiners for all classical labels, see \Ref{GIcs2}.

We close with some possible applications and open questions for future work.
Just like the HKCS, the TGCS are kinematical. They can nonetheless be used to define effective Hamiltonians (for instance, around a homogeneous background for cosmological applications, or for perturbations around flat connections for spinfoam amplitudes). Comparing the effective dynamics obtained with two different choices of kinematical coherent states can help assess the strength of the predictions, and better control the meaning of the computed quantum corrections.
The new states can also be used to construct new quantum operators based on the so-called $P$ or $Q$ representations \cite{IoPoly,Alesci:2015yqa}.

Among the open questions, two lines of future investigation clearly stand out. First, investigating whether it is possible to find a 
holomorphic representation associated with the twisted geometry parametrization, something not achieved by the TGCS presented here.
Second, the HKCS can be derived from a graph-restriction of a continuous definition on the full LQG Hilbert space $L[{\cal A},d\mu_{\rm AL}]$, see e.g. discussion in \cite{Flori:2008nw}. This is possible thanks to the distributional interpretation of the labels. The twisted geometry parametrization is on the other hand inherently discrete, hence it is not obvious what would be a continuous definition of the coherent states that reduce to the TGCS on a fixed graph. In this respect, the path for a continuous interpretation of twisted geometries recently presented in \cite{Freidel:2020ayo} offers a new way to think about this problem. We leave these questions for future work.

\subsection*{Acknowledgments}

SiS thanks Thomas Thiemann, Andrea Dapor and Param Singh for useful discussions on  heat-kernel coherent states.

\appendix
\setcounter{equation}{0}
\renewcommand{\theequation}{\Alph{section}.\arabic{equation}}

\section{Conventions}\label{AppA}

\subsection{Twisted geometries and spinorial parametrization}

On $T^*\SU(2)\simeq \SU(2)\times\su(2)$, we define the left-invariant and right-invariant vector fields acting respectively as right and left derivatives,
\begin{equation}\label{RLder}
L_if(g):=i\nabla_i^R f(g) = i\frac{d}{ds}f\left.\left(ge^{sX}\right)\right|_{s=0}, 
\qquad R_if(g):=i\nabla_i^L f(g) = i\frac{d}{ds}f\left.\left(e^{-sXg}\right)\right|_{t=0}. 
\end{equation}
They are related by the adjoint action
\be
L = -g R g^{-1},
\ee
where we used the map $L^A{}_B:=\t_i^A{}_B L^i$, and $\t_i=(-i/2)\s_i$ in terms of Pauli matrices.
The Poisson structure is defined by the following brackets,
\begin{align}\label{TSU2PB}
& \{L^i,L^j\}=\epsilon_{ijk}L_k, && \{R^i,R^j\}=\epsilon^{ijk}R^k, &&   \{L^i,R^j\}=0,\\\nn
& \{L^i,g^A{}_B\}= (g\t^i)^A{}_B, &&  \{R^i,g^A{}_B\}=-(\t^ig)^A{}_B, && \{g^A{}_B, g^C{}_D\}=0.
\end{align}
For reference, the relation to the tilde/untilde notation used in \cite{twigeo} is $X=R$,\, $\tilde{X}=L$.

The spinorial parametrization  \cite{twigeo2,Livine:2011gp,IoWolfgang}  is based on the isotropic reduction $\C^4/\!/C\simeq T^*\SU(2)$, with $(z^A_\sL,z^A_\sR)\in\C^4$, $A=0,1$, and 
$C:=\norm{z_\sL}^2-\norm{z_\sR}^2$. Here $\norm{z}^2:=\d_{A\dot A}\bar z^{\dot A} z^A $ is the SU(2)-invariant hermitian norm. Spinorial indices are raised and lowered as $z_A:= z^B\eps_{BA}$, with $\eps^{AB}=i\s_2=\eps_{AB}$ as matrices. We will also use the ket notation $\ket{z}:=z^A$, $\bra z:=\d_{A\dot A}\bar z^{\dot A}$, $\norm{z}^2=\bra{z}z\ra$, and $[z|:=z_A$, $|z]=\d^{A\dot A}\bar z_{\dot A}$. The isotropic reduction is given by \cite{twigeo2}
\be\label{spinorial}
\vec L:=\bra{z_\sL}\f{\vec \s}2\ket{z_\sL}, \qquad \vec R:=- \bra{z_\sR}\f{\vec\s}2\ket{z_\sR}, \qquad 
g :=\f{\ket{z_\sL} \bra{  z_\sR}+ |z_{\sL}] [z_{\sR}| }{\norm{z_\sL}\, \norm{z_\sR}},
\ee
with brackets
\be
\{z_\sL^A,\bar z_\sL^{\dot A} \} = -i\d^{A\dot A}, \qquad \{z_\sR^A,\bar z_\sR^{\dot A} \} = i\d^{A\dot A}.
\ee

Twisted geometries are based on the alternative parametrization of the holonomy-flux algebra in terms of both left and right-invariant vector fields, plus an additional twist angle $\xi$ that captures the remaining component of the holonomy \cite{twigeo}. This parametrization depends on a choice of section for the Hopf bundle $S^3\simeq(S^2,S^1)$. Choosing the section 
\be\label{HopfSection}
n(\z):=\f1{\sqrt{1+|\z|^2}} \mat{1}{\z}{-\bar\z}{1},\qquad \z\in\C P^1,
\ee
we have \Ref{twigeo} in the main text, here reported for convenience:
\begin{align}
& (g,\vec L) \mapsto (A,\xi, \z_{\sL}, \z_{\sR}), \qquad A\geq 0, \quad \xi\in[-2\pi,2\pi), \quad \z_\sL\in\C P^1,  \quad \z_\sR\in\C P^1, 
\\\label{twigeoApp}
& g=n_{\sL}e^{\xi\t_3}n_{\sR}^{-1}, \qquad L = An_\sL \t_3 n_\sL^{-1}, \qquad R= -An_\sR \t_3 n_\sR^{-1}. 
\end{align}
The $4\pi$ period of the twist angle can be easily identified noticing that the Hopf section does not include the $\SU(2)$ element $-\Id$.
An interesting property of this parametrization is that the twist angle is canonically conjugate to the area  under the holonomy-flux algebra \Ref{TSU2PB},
\be
\{\xi,j\}=1.
\ee
See \cite{twigeo} for more details.

The relation to the spinorial parametrization is obtained identifying the spinor components as homogeneous coordinates for the sphere $S^2\simeq \C P^1$ with stereographic coordinate $\z$. 
Choosing $\z:=-\bar z^1/\bar z^0$, we have
\begin{align}
A = \f{\norm{z_\sL}^2}2 = \f{\norm{z_\sR}^2}2,\qquad \z_\sL:=-\f{\bar z^1_\sL}{\bar z^0_\sL}, \qquad \z_\sR:=-\f{\bar z^1_\sR}{\bar z^0_\sR}, \qquad \xi=2\arg z_\sR^0-2\arg z_\sL^0.
\end{align}
One can explicitly check that the parametrizations of the vector fields in \Ref{twigeoApp} and \Ref{spinorial} are consistent. 
In view of the quantization, it is useful to consider the exponential
\be\label{defexi}
e^{i\xi} = \f{(\bar  z_\sL^0)^2}{|z_\sL^0|^2} \f{(z^0_\sR)^2}{|z_\sR^0|^2}, 
\ee
which satisfies
\be\label{exijPB}
\{e^{i\xi},A\} = ie^{i\xi}.
\ee

\subsection{Holonomy-flux algebra}

We take units $\hbar=1=G$, and use the quantization map $[,]=i\widehat{\{,\}}$ for the fundamental operators. The quantum holonomy-flux algebra reads
\begin{align}\label{HFalgebra}
& [\hat L^i,\hat L^j] = i\eps^{ijk}\hat L^k, \qquad [\hat R^i,\hat R^j] = i\eps^{ijk}\hat R^k,\qquad [\hat L^i,\hat R^j]=0, \\
& [\hat L^i, \hat g^A{}_B] = i(\hat g\t^i)^A{}_B, \qquad [\hat R^i, \hat g^A{}_B] = -i(\t^i\hat g)^A{}_B, \qquad [\hat g^A{}_B,\hat g^C{}_D]=0. 
\end{align}
The space $L_2[\SU(2),d\m_{\rm Haar}]$ carries a unitary irreducible representation of this algebra, with the flux operators acting as the left and right derivatives \Ref{RLder}, and the holonomy operator  $\hat g^A{}_B=g^A{}_B$ acting multiplicatively.
The eigenvectors $\ket{g}$ of the holonomy operator provide a generalized basis (i.e. plane-wave-like), and the eigenvectors $\ket{j,m,n}$ of $\hat{\vec L}^2,\hat L^3,\hat R^3$ an orthogonal basis.
 The transformation between the two is given by the Wigner matrices,
\be
\bra{\bar g}j,m,n\ra = \sqrt{d_j} D^{(j)}_{mn}(g) = \sqrt{d_j} \bra{j,m}g\ket{j,n}.
\ee
The expression of this overlap invites also the convenient interpretation $\ket{j,m,n}\mapsto\sqrt{d_j}\ket{j,n}\bra{j,m}$, in line with the Peter-Weyl decomposition $L_2[\SU(2),d\m_{\rm Haar}]=\oplus (V^{j}\otimes\bar V^{j})$.\footnote{Care is however needed in using this interpretation, since these vectors live in different Hilbert spaces. In particular, $\la{k,p,q}\ket{j,m,n}=\d_{kj}\d_{pm}\d_{qn}\neq d_j \d_{kj}\d_{pm}\d_{qn}$.}
Accordingly, the right-invariant vector fields act as hermitian conjugated. 
With our conventions, the action of the vector fields is
\begin{align}
& \hat L_3 \ket{j,m,n}\ket{j,m,n} = n\ket{j,m,n}, &&  \hat R_3 \ket{j,m,n}\ket{j,m,n} = -m\ket{j,m,n}, \\\nn
& \hat L_\pm \ket{j,m,n}\ket{j,m,n} = c_{\pm}(j,n)\ket{j,m,n\pm 1}, && \hat R_\pm \ket{j,m,n}\ket{j,m,n} = -c_{\mp}(j,m)\ket{j,m\mp 1,n},
\end{align}
where $c_{\pm}(j,m):=\sqrt{(j\mp m)(j\pm m+1)}$. The minus sign in the action of the right-invariant operators follows from the relative sign in \Ref{RLder}, and it is chosen to have the structure constants of both sets of operators with the same sign, see \Ref{HFalgebra}. The action of the holonomy operator is obtained through the Clebsch-Gordan coefficients $C^{j1j2jJ}_{m_1m_2M}$ as
\be
\hat g^A{}_B\ket{j,m,n} = \sum_{M,N} C^{\tfrac12,j,j+\tfrac12}_{A,m,M} C^{\tfrac12,j,j+\tfrac12}_{B,n,N} \ket{j+\tfrac12,M,N}
+\sum_{M,N} C^{\tfrac12,j,j-\tfrac12}_{A,m,M} C^{\tfrac12,j,j-\tfrac12}_{B,n,N} \ket{j-\tfrac12,M,N}
\ee
It includes both positive and negative shifts of the spins, and generic mixtures with shifts of the magnetic numbers.

Following Schwinger, these actions can be realized in terms of two pairs of harmonic oscillators, $a_\sL^A$ and $a_\sR^A$, with $A=0,1$. To take into account the hermitian action of the right-handed generators, we interpret
\be
\ket{j,m,n} = \sqrt{d_j}\ket{n_\sL^0,n_\sL^1}\bra{n_\sR^0,n_\sR^1},
\ee
and identify
\begin{equation}
j = \frac{n_\sL^0+n_\sL^1}{2}=\frac{n_\sR^0+{n}_\sR^1}{2}, \qquad m = \frac{n_\sR^0-n_\sR^1}{2},
\qquad  n = \frac{{n}_\sL^0-{n}_\sL^1}{2}.
\label{id1}
\end{equation}
The expression for the generators is
\begin{align}
 \hat{L}^{\pm,3} = \left(a_\sL^{0\dagger}{a_\sL}^1,\quad {a_\sL}^0 {a_\sL}^{1\dagger},\quad \frac{{n_\sL}^0-{n_\sL}^1}{2}\right),
\qquad \hat{R}^{\pm,3}= -\left(a_\sR^{0\dagger}a_\sR^1,\quad a_\sR^0a_\sR^{1\dagger},\quad \frac{n_\sR^0-n_\sR^1}{2}\right).
\end{align}
These can be obtained from the spinorial representation \Ref{spinorial} and the standard Schr\"odinger harmonic oscillator quantization map
\be\label{ztoa}
z \mapsto \hat z:= a, \qquad \bar z\mapsto \hat{\bar z}:=a^\dagger.
\ee
The resulting brackets are
\be
[a_\sL^A,a_\sL^{B\dagger}] = \d^{AB}, \qquad [a_\sR^A,a_\sR^{B\dagger}] = -\d^{AB}.
\ee

Notice that one can also give a representation of the holonomy operator in terms of harmonic oscillators, by applying the above map to the spinorial parametrization in \Ref{spinorial}. This construction requires a choice of ordering, see \cite{Livine:2013zha}.
Upon doing so, one obtains the full holonomy-flux quantum algebra in terms of harmonic oscillators, as advocated in \cite{twigeo2,FreidelUNcs,EteraSpinor}. For recent applications of this formalism to understanding the structure of correlations and entanglement on spin network states in loop quantum gravity, see \cite{Bianchi:2016hmk,Bianchi:2018fmq}.

\subsection{Shift operators}\label{AppShift}

A further point of interest of the twisted geometries and spinorial parametrizations is to permit the construction of new operators. This is similar to the way the Schwinger representation for the algebra permits the construction of new operators for the angular momentum and its invariants \cite{Girelli:2005ii,EteraSU2UN,IoHolo}. In the case of the holonomy-flux algebra, one can define new operators to simplify the action
of the holonomy operator in the electric basis, providing for instance only positive or only negative shifts, as well as disentangling the shifts on the spins from the shifts on the magnetic numbers. 
One such new operator is the twist angle, which is canonically conjugated to the area, and thus expected to provide a spin shift in one direction only. We can define it from its classical expression \Ref{defexi} and the quantization map \Ref{ztoa}. With a natural ordering prescription, we define
\be\label{defexi}
\exiop := (n_\sR^0)^{-1} (a_\sR^{0\dagger})^2 (n_\sL^0)^{-1} (a_\sL^0)^2.
\ee
We notice that 
\be
[\exiop,\hat \jmath] = -\exiop,
\ee
as to be expected for a creation operator, and in agreement with \Ref{exijPB}. Its action can be easily computed to be
\be
\exiop \ket{j,m,n} = \f{(j+m+1)^{1/2}}{(j+m+2)^{1/2}} \f{(j+n+1)^{1/2}}{(j+n+2)^{1/2}} \ket{j+1,m+1,n+1}.
\ee

This operator is sufficient to the scopes of our paper. We used it in the main text to evaluate expectation values of holonomy components on the heat-kernel and twisted geometry coherent states, and compare them. 
To define the inverse of \Ref{defexi}, a bit more care is required: merely taking the adjoint leads to an operator mapping $\ket{j,m,n}\mapsto\ket{j-1,m-1,n-1}$ unboundedly. A similar problem appears when using action-angle operators for the harmonic oscillator, and can be dealt with redefining the creation and annihilation operators as $\hat \jmath \exiop$ and $\widehat{e^{-i\xi}}\hat \jmath$, so to have a bounded action for $\ket{0,0,0}$. See e.g. \cite{Leacock87}.

We remark also that additional useful operators that perform different shifts. For instance it is possible to shift the spin and not the magnetic numbers,
\begin{align}
& \D_j:= a_\sR^{0\dagger} a_\sR^{1\dagger} a_\sL^{0} a_\sL^{1}, \\
& \D_j\ket{j,m,n} = \sqrt{(j+m)(j-m)(j+n)(j-n)} \ket{j-1,m,n}.
\end{align}
This action vanishes on any highest or lowest magnetic weight, as well as on the vacuum ket $\ket{0,0,0}$, making it well-defined on $L_2[\SU(2),d\m_{\rm Haar}]$. The hermitian conjugate gives
\be
\D_j^\dagger \ket{j,m,n} = \sqrt{(j+m+1)(j-m+1)(j+n+1)(j-n+1)} \ket{j+1,m,n}.
\ee

We conclude with a list of properties of Wigner's matrices that were used in the main text.
\be\label{WignerHighest}
D^{(j)}_{m,-j}(g)=\binom{2j}{j+m}^{1/2} (g_{12})^{j+m} (g_{22})^{j-m}, \qquad
D^{(j)}_{mj}(g)=\binom{2j}{j-m}^{1/2} (-\bar g_{12})^{j-m} (\bar g_{22})^{j+m}.
\ee
\be
D^{(j)}_{m,-j}(n)
=
\sqrt{\f{(2j)!}{(j+m)!(j-m)!}}\,
\f{\zeta^{j+m}}{(1+|\zeta|^2)^j}
\,, \qquad
D^{(j)}_{m,j}(n)
=
\sqrt{\f{(2j)!}{(j+m)!(j-m)!}}\,\f{(-\bar\z)^{j-m}}{(1+|\zeta|^2)^j}.
\ee
\begin{align}
& \sum_n \left( \f{j+n+1}{j+n+2} \right)^{\tfrac12} D^{(j)}_{nj}(n_{\sL})  D^{(j+1)}_{n+1,j+1}(\bar n_{\sL})
= \f{ (1+2j-|\z_{\sL}|^2)(1+ |\z_\sL|^2)^{d_j} +|\z_\sL|^{4(j+1)}  }{\sqrt{(2j+1)(2j+2)} \, (1+|\z_{\sL}|^2)^{2j+1}}, 
\\
& \sum_n \left( \f{j+n+1}{j+n+2} \right)^{\tfrac12} D^{(j)}_{n,-j}(n_{\sL})  D^{(j+1)}_{n+1,j+1}(\bar n_{\sL})
= \f{ (-\z_\sL)^{2j} }{\sqrt{(2j+1)(2j+2)} \, (1+|\z_{\sL}|^2)^{2j+1}}, 
\\
& \sum_n \left( \f{j+n+1}{j+n+2} \right)^{\tfrac12} D^{(j)}_{nj}(n_{\sL})  D^{(j+1)}_{n+1,-j-1}(\bar n_{\sL})
= \f{ (-\bar\z_\sL)^{2j+2} }{\sqrt{(2j+1)(2j+2)} \, (1+|\z_{\sL}|^2)^{2j+1}}, 
\\
& \sum_n \left( \f{j+n+1}{j+n+2} \right)^{\tfrac12} D^{(j)}_{n,-j}(n_{\sL})  D^{(j+1)}_{n+1,-j-1}(\bar n_{\sL})
= \bar\z_\sL^2\f{ (-1+d_j|\z_{\sL}|^2)(1+ |\z_\sL|^2)^{d_j} +1  }{\sqrt{(2j+1)(2j+2)} \, (1+|\z_{\sL}|^2)^{2j+1}\, |\z_\sL|^{4}}.
\end{align}

\section{Coherent intertwiners}\label{AppCI}
The intertwiners depend on the orientations of the links. For a 3-valent node with all links outgoing, we define
\be
i_{n_3} = \Wthree{j_1}{j_2}{j_3}{m_1}{m_2}{m_3}.
\ee
For a 4-valent node, choosing for instance the recoupling channel $j_{12}$, we have
\be
i_{n_4} = \Wfour{j_1}{j_2}{j_3}{j_4}{m_1}{m_2}{m_3}{m_4}{j_{12}} = \sum_{m_{12}} (-1)^{j_{12}-m_{12}}\Wthree{j_1}{j_2}{j_{12}}{m_1}{m_2}{m_{12}}\Wthree{j_{12}}{j_3}{j_4}{-m_{12}}{m_3}{m_4},
\ee
and so on for higher valent nodes.
For ease of notation, we use $i$ to indicate both the invariant tensor and the quantum label without making the choice of recoupling channel explicit. When a link is instead incoming, the relevant intertwiner is obtained acting with $\eps^{(j)}$, where
$\eps^{(j)}_{mn}=(-1)^{j-m}\d_{m,-n}$. Similarly, also the coefficients of the coherent intertwiners \cite{LS} depend on the orientation. For a node with all links outgoing, we define
\begin{subequations}\label{coeffint}\be
c_{i}^{\scr{\downarrow\downarrow\downarrow\downarrow}}(\vec n_{i}) = \bra{i}j_i,\vec n_i\ra
:= \sum_{m_i} \Wfour{j_1}{j_2}{j_3}{j_4}{m_1}{m_2}{m_3}{m_4}{i} \prod_{i=1}^4\bra{j_i, m_i} n_{i} |j_i, -j_i \ra,
\ee
following the conventions of \cite{Dona:2017dvf}.
If the link 1 is incoming, we define
\be
c_{i}^{\scr{\uparrow\downarrow\downarrow\downarrow}}(\vec n_{i}) = \bra{i}j_i,\vec n_i\ra := 
\sum_{m_i} \Wfour{j_1}{j_2}{j_3}{j_4}{m_1}{m_2}{m_3}{m_4}{i} \bra{j_1, m_1} n_{1} |j_1, j_1 \ra \prod_{i\neq 1} \bra{j_i, m_i} n_{i} |j_i, -j_i \ra.
\ee\end{subequations}
And so on. The rule is that the coefficient for each outgoing link is computed using the lowest weight, and for each incoming using the highest weight. 
To help the memory for this rule, we denote an outgoing link with a downward arrow.

With our notation to label the Hopf section at the source and target respectively $n_{\sL}$ and $n_{\sR}$, 
group averaging a coherent state $D^{(j)}_{n,-j}(n_{\sL})D^{(j)}_{-j,m}(n_{\sR}^{-1})$ produces coherent intertwiners with $n_{\sL}$ in the lowest weight for each outgoing link, and  $n_{\sR}$  in the highest weight for each incoming link. Switching to 
$D^{(j)}_{n,j}(n_{\sL})D^{(j)}_{j,m}(n_{\sR}^{-1})$ we have  $n_{\sL}$ in the highest weight for each outgoing link, and  $n_{\sR}$  in the lowest weight for each incoming link.
In other words, the fixed orientation of the links determines uniquely whether it is $n_{\sL}$ or $n_{\sR}$ that enters the coherent intertwiner, and the form of the state (namely lowest or highest weight) determines the explicit form of the coefficient.

\providecommand{\href}[2]{#2}\begingroup\raggedright\endgroup

\end{document}